\documentclass[sigconf, nonacm]{acmart}

\newcommand\vldbdoi{XX.XX/XXX.XX}

\usepackage{amsmath,amsfonts,amsthm}
\usepackage{epsfig,xspace,url,xcolor,color}

\usepackage[linesnumbered, ruled, algo2e, noresetcount]{algorithm2e}
\usepackage{setspace} 

\usepackage{graphicx}
\usepackage{textcomp}
\usepackage[caption=false]{subfig}
\usepackage[latin1]{inputenc}
\usepackage{float}
\usepackage{placeins}
\usepackage{listings}
\usepackage[T1]{fontenc}
\usepackage{caption}
\usepackage{enumitem}
\usepackage{csquotes}
\usepackage{marginnote}
\usepackage{multirow}
\usepackage{diagbox}
\usepackage{lipsum}
\usepackage{tikz}

\newcommand\blackcircle[1]{
  \tikz[baseline=(X.base)]
    \node (X) [draw, shape=circle, inner sep=0, fill=black, text=white] {#1};
}

\newcommand\whitecircle[1]{
  \textcircled{\small{#1}}
}

\makeatletter  
\newcommand{\my@arrow}[1]{\ooalign{$#1-\mkern-8mu-$\cr\hidewidth$#1>$}}
\newcommand{\myarrow}{\mathrel{\mathpalette\my@arrow\relax}}
\makeatother

\let\oldnl\nl
\newcommand{\nonl}{\renewcommand{\nl}{\let\nl\oldnl}}

\newcommand\Paragraph[1]{\vspace{0.02in}  \noindent \textbf{#1}}
\newcommand\SubParagraph[1]{\vspace{0.02in}  \noindent \textit{\underline{\smash{#1}}}}

\newcommand{\vn}[1]{\textsf{#1}}
\newcommand{\fn}[1]{\textsc{#1}}

\newcommand{\sn}{\fn{CUBIT}\xspace}
\newcommand{\sns}{\fn{CUBIT}} 

\newcommand{\ub}{UpBit\xspace} 
\newcommand{\ubs}{UpBit} 

\newcommand{\bt}{B$^+$-Tree\xspace} 
\newcommand{\bts}{B$^+$-Tree} 

\begin{document}
\title[\fn{\sns}: Concurrent Updatable Bitmap Indexing (Extended Version)]{\fn{\sns}: Concurrent Updatable Bitmap Indexing\\(Extended Version)}

\author{Junchang Wang}
\affiliation{\institution{Nanjing University of Posts and Telecommunications}\country{}}
\email{wangjc@njupt.edu.cn}

\author{Manos Athanassoulis}
\affiliation{\institution{Boston University}\country{}}
\email{mathan@bu.edu}

\begin{abstract}
     Bitmap indexes are widely used for read-intensive analytical workloads because they are clustered and offer efficient reads with a small memory footprint.
However, they are notoriously inefficient to update.
As analytical applications are increasingly fused with transactional applications, leading to the emergence of hybrid transactional/analytical processing (HTAP), it is desirable that bitmap indexes support efficient concurrent real-time updates.
In this paper, we propose \textbf{C}oncurrent \textbf{U}pdatable \textbf{Bit}map indexing (\sns) that offers efficient real-time updates that scale with the number of CPU cores used and do not interfere with queries.
Our design relies on three principles.
First, we employ a horizontal bitwise representation of updated bits, which enables efficient atomic updates without locking entire bitvectors.
Second, we propose a lightweight snapshotting mechanism that allows queries (including range queries) to run on separate snapshots and provides a wait-free progress guarantee.
Third, we consolidate updates in a latch-free manner, providing a strong progress guarantee.
Our evaluation shows that \sn offers 3--16$\times$ higher throughput and 3--220$\times$ lower latency than state-of-the-art updatable bitmap indexes. 
\sns's update-friendly nature widens the applicability of bitmap indexing.
Experimenting with OLAP workloads with standard, batched updates shows that \sn overcomes the maintenance downtime and outperforms DuckDB by 1.2--2.7$\times$ on TPC-H.
For HTAP workloads with real-time updates, \sn achieves 2--11$\times$ performance improvement over the state-of-the-art approaches.

\end{abstract}

\maketitle



\section{Introduction}\label{sec.intr}

\Paragraph{Access Path Selection.}
When a query targets columns with secondary indexes, Database Management Systems (DBMSs) evaluate the available \textit{access paths} based on characteristics like selectivity to choose between an index scan and a full sequential scan \cite{Kester2017}.
Historically, indexes are tree-based structures like B$^+$-trees \cite{Lehman1981, Levandoski2013, Wang2018a} and tries \cite{Leis2013, Mao2012}.
It is widely accepted that trees are valuable only for extremely selective queries; otherwise, scans are better \cite{Silberschatz2020, Graefe2011}.

\Paragraph{Bitmap Indexes.}
For read-only workloads, bitmap indexes are a great alternative, especially when each indexed attribute has a moderate number of $c$ unique values (\emph{cardinality}) \cite{Silberschatz2020}.
In its basic form, each bitmap index targets a specific attribute \cite{Chan1998}. 
In the most common design, termed \emph{equality encoding}, each value of the attribute's domain is associated with a \emph{bitvector}, which contains $n$ bits, where $n$ is the number of tuples in the indexed table.
The conceptual $c \times n$ bit matrix that the bitvectors form allows queries to be quickly answered using efficient bitwise instructions \cite{Chan1998, Wu2009, Athanassoulis2016a}.

\begin{figure}[t]
    \centering
    \setlength{\belowcaptionskip}{-15pt}

    \includegraphics[width=0.92\columnwidth]{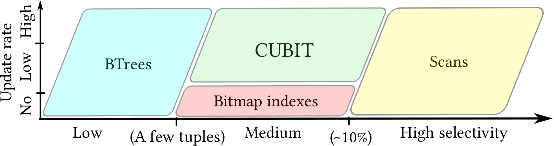}
    \label{eva.motivation.designSpace}

    \vspace{-0.1in}
    \caption{In the presence of both queries and updates, access path selection between tree-based indexing, bitmap indexing, and sequential scans depends on selectivity and update rate. Unlike prior bitmap indexes that were mainly used for read-only workloads, our solution, \sns, enables bitmap indexing for higher update rates: OLAP with batched updates (\S\ref{sec.eva.olap}) and HTAP with real-time updates (\S\ref{sec.eva.htap}).}
	\label{fig.motivation}
\end{figure}

\Paragraph{Bitmap Index as a Secondary Index.}
Besides their fast query performance, bitmap indexes offer intrinsic benefits when used as secondary indexes in DBMSs.
First, they have a \emph{small memory footprint} because, apart from the bit-matrix and a mapping from the key domain to the corresponding bitvectors (columns of the bit-matrix), they do not store extra metadata like keys and pointers to tuples.
Second, they are \textit{clustered} in the sense that the order of the entries in the generated bitvector follows the physical order of the tuples.
This significantly boosts query performance over traditional secondary indexes, which are generally unclustered.
Third, bitmap indexes are \emph{composition-friendly}, that is, multiple bitmap indexes on different attributes can be composed on demand, which outperforms multi-column tree indexes for large datasets (more details in  \S\ref{sec.bitmap.merits}).
As an example, using a row-based DBMS for analytical queries, our bitmap index reduces memory consumption by 92\% and boosts query performance by 1.7--2.6$\times$ when compared with tree-based multi-column indexes (more details in \S\ref{sec.eva.olap.dbx1000}).
In addition, for moderate selectivity, bitmap indexes incur less CPU cache and TLB misses than scans.
For example, DuckDB~\cite{Raasveldt2019} (a column-based OLAP DBMS) with our bitmap index achieves a $\sim$2$\times$ performance improvement over its highly optimized scan, and this trend continues until the selectivity reaches up to 10\% (\S\ref{sec.eva.olap.duckdb}).

\Paragraph{Challenges with Bitmap Indexes.}
Despite their benefits, bitmap indexes are notoriously expensive to update.
In order to support updates, DBMSs either drop existing bitmap indexes and then rebuild them from scratch~\cite{Stockinger2009, Li2020a}, 
or perform updates in a batch mode and in a pre-defined time period, during which the index is unavailable for queries \cite{Burleson2011}, trading off system availability for fast queries.
Thus, \textit{bitmap indexes are rarely used for workloads with updates}, as illustrated in Figure \ref{fig.motivation}.

To address this challenge, recently proposed \emph{updatable bitmap indexes}~\cite{Athanassoulis2016a, Canahuate2007} support real-time \vn{Update}, \vn{Delete}, and \vn{Insert} operations (henceforth, \emph{UDI}s in short) using additional bitvectors to capture incoming updates in an \textit{out-of-place} manner.
In particular, UCB \cite{Canahuate2007} uses a single bitvector to invalidate rows, while UpBit \cite{Athanassoulis2016a}, the state-of-the-art solution, associates one new bitvector with each value of the domain to serve as a buffer for UDIs (\S\ref{sec.bitmap.bitmap}).

When integrating these approaches into modern DBMSs, we found the following three drawbacks.
(1) Prior updatable bitmap indexes were designed to run sequentially.
When used on multicores, in-progress UDIs block concurrent queries that may fail to meet performance requirements (e.g., from a Service-Level Agreement).
(2) UDIs on bitmap indexes flip bits in bitvectors that are commonly compressed using Run-Length-Encoding-based techniques like WAH \cite{Wu2006}.
To do this, one has to decode each bitvector, flip the corresponding bits, and re-encode.
This \emph{decode-flip-encode} procedure is time-consuming with large datasets.
(3) Skewed datasets raise even higher contention around a few hot-spot bitvectors.

\begin{figure}[t]
    \centering
    \setlength{\belowcaptionskip}{-10pt}
    \includegraphics[width=1\columnwidth]{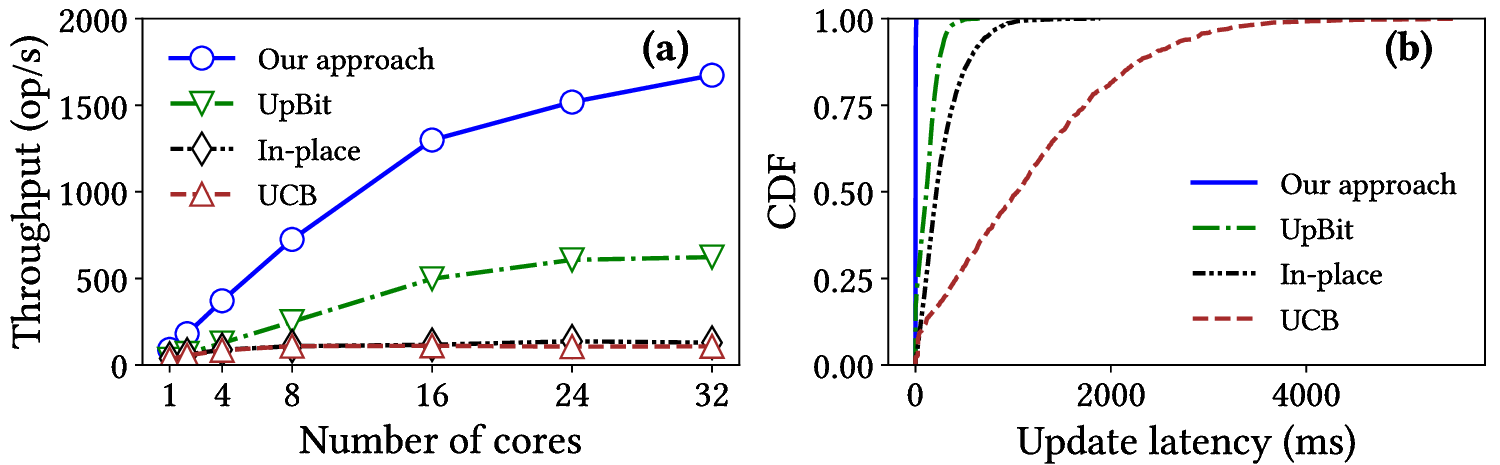}
    \vspace{-0.23in}
    \caption{(a) Existing updatable bitmap indexes do not scale on multicores,
            and (b) their synchronization mechanisms incur long tail latency.}
	\label{fig.motivation.scale}
    \vspace{-0.1in}
\end{figure}

To demonstrate these performance issues, we parallelize and experiment with three bitmap index designs that support updates, \emph{In-place}~\cite{Silberschatz2020}, \emph{UCB}~\cite{Canahuate2007}, and \emph{UpBit}~\cite{Athanassoulis2016a}, by using a 48-core system (see \S\ref{sec.eva} for experimental setup).
Figure \ref{fig.motivation.scale}a shows that In-place and UCB scale only up to 4 CPU cores, while \ub scales better but plateaus at 16 cores.
Moreover, because of the increased contention at high concurrency levels, these bitmap indexes suffer from surprisingly long tail latency.
Figure \ref{fig.motivation.scale}b shows that with 16 cores, their UDI latency can reach up to 6 seconds (see \S\ref{sec.eva.scale} for queries). 

\Paragraph{Our Approach.}
To solve the aforementioned drawbacks, we propose \sns, a new design for \textbf{C}oncurrent \textbf{U}pdatable \textbf{Bit}map indexing.
\sn offers real-time updates and scales with the number of CPU cores, as shown in Figure \ref{fig.motivation.scale} (blue lines).

From a bird's-eye view, \sn performs UDIs on compressed value bitvectors (VB) in an out-of-place manner, leverages multi-versioning for wait-free queries \cite{Herlihy1991}, and adopts latch-free techniques for real-time UDIs.
Overall, \sn's design encompasses:

\SubParagraph{(A) Horizontal Update Deltas (HUD).} 
We introduce HUD, a horizontal representation of the bits flipped by each UDI (\S\ref{sec.overview.hud}).
Contrary to prior work that buffers UDIs in additional bitvectors, we organize out-of-place updates per row of the bit-matrix in contiguous memory, 
concentrating all bit-flips to a single place (instead of multiple bitvectors). 
HUDs enable atomic UDIs without locking entire VBs, resolving the major update bottleneck for bitmap indexes (\S\ref{sec.challenge}). 

\SubParagraph{(B) Lightweight Snapshots.} 
We develop a lightweight snapshotting mechanism for bitmap indexes (\S\ref{sec.overview.mvcc}) using compact deltas (\S\ref{sec.overview.deltalog}), batched merges (\S\ref{sec.overview.merge}), and segmentation (\S\ref{sec.overview.segbtv}).
\sn allows queries and UDIs to work on \textit{different index snapshots} and guarantees that an analytical query can be completed in a finite number of steps, even when UDIs are in progress.
In contrast, the atomicity granularity of tree-based indexes (e.g., Bw-Tree \cite{Levandoski2013} and ART \cite{Leis2013}) is smaller, and their range queries could be interrupted by UDIs.

\SubParagraph{(C) Scalable Synchronization.}
Concurrency control mechanisms like two-phase locking (2PL) and multi-version concurrency control (MVCC) use latches to serialize UDIs on the same portion of data.
However, updatable bitmap indexes face high contention because (1) UDIs lock bitvectors (which are typically significantly fewer than records or pages that are locked when using other indexes), (2) and skewed UDIs concentrate around a few hot-spots bitvectors leading to even higher contention (further discussed in \S\ref{sec.challenge}).
To meet the time constraints for indexing, we address this issue by developing a \emph{consolidation-aware} \cite{Johnson2010} and \emph{latch-free} \cite{Herlihy1990} UDI mechanism (\S\ref{sec.overview.sync}).
Rather than competing with each other, \sns's UDIs work synergistically to reduce contention, thus making \sn scalable even with high UDI ratios and skewed datasets.

\Paragraph{Broader Applicability.}
Being update-friendly, \sn widens the applicability of bitmap indexing by enabling DBMSs to maintain bitmap indexes on frequently updated attributes, in a variety of use cases, including OLAP with batched updates (\S\ref{sec.eva.olap}) and HTAP with real-time updates (\S\ref{sec.eva.htap}), the green area shown in Figure \ref{fig.motivation}.
\sn instances not only reduce the amount of data read from storage for the Scan operator, but also provide sufficient information for the Join and Aggregation operators, eliminating the need to build the costly intermediate data structures (e.g., hash table for Joins).

\Paragraph{Contributions.}
Overall, we make the following contributions.

\begin{itemize}[leftmargin=1.2em, itemsep=0.22em] 

    \item We parallelize state-of-the-art bitmap indexes and analyze their bottlenecks.
        Based on the gained insights, we propose \sns, the first concurrent updatable bitmap index.

    \item We extensively evaluate \sn using synthetic workloads and industry-grade benchmarks, showing that it offers 3--16$\times$ higher throughput than the baselines at different concurrency levels, with 4--13$\times$ lower query latency and 3--220$\times$ lower UDI latency.

    \item \sn expands the applicability of bitmap indexing by enabling, for the first time, bitmap indexes on frequently updated attributes.
    We demonstrate this by integrating \sn in both DBx1000~\cite{Yu2014} and DuckDB~\cite{Raasveldt2019}, using both OLAP and HTAP workloads.

\begin{itemize}[leftmargin=1.2em, itemsep=0.22em] 

    \item Our evaluation in DBx1000 on TPC-H with standard refresh operations~\cite{TPCH} shows that \sn (1) does not introduce any maintenance overhead, and (2) compared to state-of-the-art indexes (ART, Bw-Tree, and \bt), provides 1.7--2.2$\times$ higher throughput with up to 92\% lower memory footprint.

    \item Our evaluation shows that the \sns-powered query engine achieves a 1.2--2.7$\times$ query performance improvement on 12 out of 22 TPC-H queries compared to optimized native approaches in DuckDB.
    
    \item Experimenting with CH-benCHmark~\cite{Cole2011} with real-time updates shows that \sn provides 2.1--11.2$\times$ higher throughput than DBx1000 (with indexes and scans).

\end{itemize}

    \item Overall, our results show that \sn is a promising indexing candidate for selective queries on workloads with updates.

\end{itemize} 

\section{Background on Bitmap Indexes}\label{sec.bitmap}

\label{sec.bitmap.bitmap}

\begin{figure}[t]
    \centering
    \setlength{\belowcaptionskip}{-10pt}
    \includegraphics[width=0.9\columnwidth]{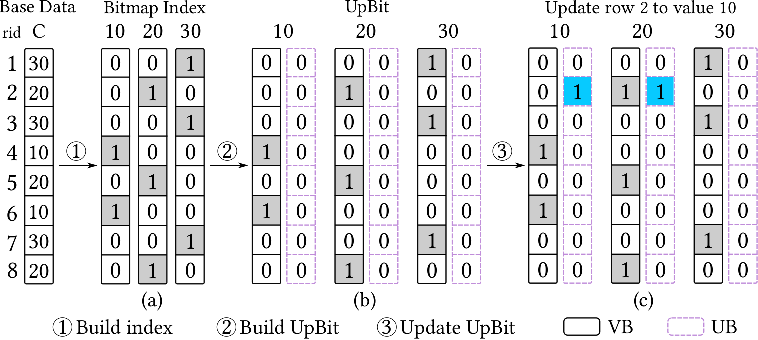}
    \vspace{-0.1in}
    \caption{
        (a) A classic bitmap index.
        (b) The state-of-the-art updatable bitmap index \ub~\cite{Athanassoulis2016a} that associates a UB with each VB.
        (c) \ubs's UDIs update highly-compressible UBs.}
	\label{fig.bitmap}
\end{figure}

\Paragraph{Bitmap Indexes Basics.}
Modern bitmap index designs~\cite{ONeil1987, ONeil1997, Wu2009} introduce a \emph{bit-matrix} structure consisting of several bitvectors, one for each distinct domain value.
The $k^{th}$ bit of each bitvector is set to 1 if the $k^{th}$ row of the attribute is equal to the corresponding value; otherwise, it is set to 0.
Figure \ref{fig.bitmap}a shows an uncompressed bitmap index for an attribute with three different values.
This bit-matrix structure fits static analytical workloads; an equality query simply reads the corresponding bitvector, and a range query performs a bitwise OR between the corresponding bitvectors.

\SubParagraph{Keeping Bitvectors Small.}
Bitvectors contain a lot of 0s, thus being amenable to compression.
One class of compressing techniques uses Run-Length Encoding (RLE)~\cite{Antoshenkov1995, Wu2006, Colantonio2010, Deliege2010, Fusco2010}.
In particular, Word-Aligned Hybrid (WAH)~\cite{Wu2006} splits the original bit-string into 31-bit words.
A compressed bitvector contains two types of words: \textit{fill} and \textit{literal words}.
The first encodes long sequences of 0s or 1s using RLE.
The latter is for segments blended with 0s and 1s.
For each word, the most significant bit is used to indicate its type (\textit{fill} or \textit{literal}).
A literal word uses the remaining 31 bits to store the original bit-pattern, while a fill word uses the second most significant bit to indicate if the fill is all 0s or 1s, and the remaining bits to keep track of the number of consecutive fill words.
In our work, we use the widely used open-source implementation of WAH \cite{Wu2009, Canahuate2007, Athanassoulis2016a}.

\SubParagraph{Updatable Bitmap Indexes.}
Only a handful of bitmap indexes are update-friendly because of the costly \emph{decode-flip-encode} procedure \cite{Silberschatz2020, Canahuate2007, Athanassoulis2016a}.
The most straightforward approach, denoted \emph{In-place} \cite{Silberschatz2020}, directly updates the underlying bit-matrix.
For example, to update the $n^{th}$ row from value $v_1$ to $v_2$, \textit{In-place} decodes-flips-encodes both bitvectors for $v_1$ and $v_2$.
To reduce this cost, \emph{UCB} \cite{Canahuate2007} introduces an additional compressed \emph{existence bitvector} (EB) that indicates the validity of each row.
Initially, all bits in EB are set to 1s.
A delete operation sets the corresponding bit in EB to 0.
An update operation appends the new value to the tail of the bitvector and maps the old row ID to the new one.
The efficiency of UCB is predicated on EB being highly compressible and value bitvectors being immutable.
However, their performance deteriorates sharply as UDIs accumulate and EB becomes less compressible \cite{Athanassoulis2016a}.

\SubParagraph{\ubs.}
To address the above challenges, \ubs~\cite{Athanassoulis2016a} maintains for every value in the domain, $val$, a value bitvector {(VB)} and an extra \emph{update bitvector} {(UB)} to keep track of updates to VB (Figure \ref{fig.bitmap}b).
Both VBs and UBs are compressed.
To update the $2^{nd}$ row from value 20 to 10, \ub flips the $2^{nd}$ bits of the UBs of values 10 and 20 (Figure \ref{fig.bitmap}c).
Similarly, in order to delete the $n^{th}$ row, \ub retrieves the current value of this row and then flips the $n^{th}$ bit of the corresponding UB.
In order to insert a new entry, \ub appends 1 at the tail of the corresponding UB and increments the global variable \emph{N\_ROWS} \cite{Athanassoulis2016a}.
The core idea of \ub is to perform UDIs on UBs.
Because UBs are sparse and, thus, highly compressible, updating them is inexpensive. In several cases, it can work directly on compressed bitvectors, and when this is not possible, the decode-flip-encode cycle is lightweight.
A query on value $val$ performs a bitwise \emph{XOR} on the corresponding <VB, UB> pair to retrieve the up-to-date bitvector, 
and a range query performs bitwise \emph{OR} among the resulting bitvectors.
As 1s in UBs accumulate, \ub merges <VB, UB> pairs opportunistically (at query time) and generates new versions of VBs along with empty, highly compressible UBs~\cite{Athanassoulis2016a}.

\label{sec.bitmap.merits}

\begin{figure}[t]
    \centering
    \setlength{\belowcaptionskip}{-10pt}
    \includegraphics[width=0.82\columnwidth]{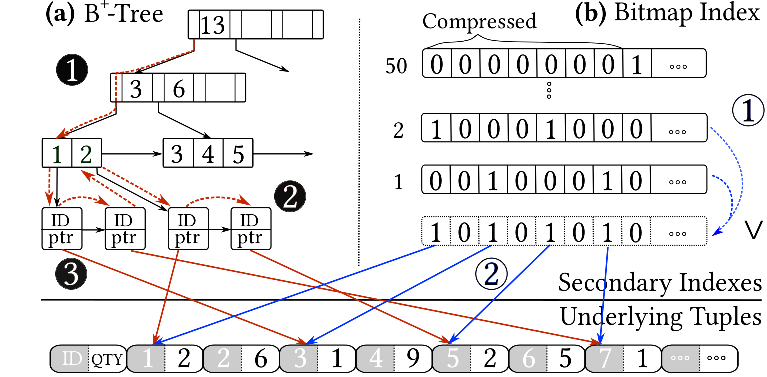}
    \vspace{-0.1in}
    \caption{Queries of the form ``\emph{QTY < 3}'' by using either a B$^+$-tree or a bitmap index. The bitmap index is clustered, composition-friendly, and space-efficient.}
	\label{fig.motivation.index}
 \vspace{-0.2in}
\end{figure}

\Paragraph{Bitmap Indexes vs. Tree Indexes.}
To motivate our goal of widening the applicability of bitmap indexes, we now compare them with tree-based indexes, as shown in Figure~\ref{fig.motivation.index}.
Consider a table with the primary-key attribute \emph{ID} and a non-primary-key attribute \emph{Quantity} (QTY for short) that spans the range of integers [1, 50] and that the underlying tuples are fixed-length (at the bottom of the figure).
A B$^+$-tree secondary index on attribute \emph{QTY} can accelerate range queries as shown in Figure \ref{fig.motivation.index}a.
We use the QTY values as the index keys and associate each leaf pointer with a linked list in case of duplicates. 
Each node in the list contains a pointer to the corresponding tuple and the \emph{ID} to keep the list ordered.
For a range query of the form \emph{``QTY < 3''}, we \blackcircle{1} traverse down the tree to the leaves, \blackcircle{2} visit the linked lists, and \blackcircle{3} access underlying tuples (red arrows in Figure \ref{fig.motivation.index}), avoiding full scans.

On the other hand, Figure \ref{fig.motivation.index}b illustrates a bitmap index on the same attribute.
The bitmap consists of 50 compressed bitvectors, which, for ease of presentation, are laid out horizontally.
In order to answer the same query \emph{``QTY < 3''}, we \whitecircle{1} perform a bitwise \emph{OR} operation between the two bitvectors of values 1 and 2 to generate an on-the-fly resulting bitvector, and \whitecircle{2} access the underlying fixed-length tuples sequentially in one pass (blue arrows).

\Paragraph{The Benefits of Bitmap Indexing} shown in Figure \ref{fig.motivation.index} are:

\SubParagraph{Clustered.}
Bitmap indexes are clustered \cite{Silberschatz2020} in the sense that the order of the entries in the generated bitvector follows the physical order of the tuples (blue arrows in Figure~\ref{fig.motivation.index}).
Therefore, queries on bitmap indexes read data pages sequentially and only once.
This addresses one significant performance drawback of existing secondary indexes that are typically unclustered (red arrows).

\SubParagraph{Small Memory Footprint.}
Being clustered, a bitmap index can efficiently transform the 1s in the resulting bitvector to row IDs, thus minimizing metadata size.
The memory consumption of a bitmap index is mainly due to its bitvectors, which can be highly compressed (\S\ref{sec.bitmap.bitmap}).
This addresses a significant space overhead of existing secondary indexes: tree nodes store metadata like pointers to tuples and, in some cases, primary key values, as shown in Figure \ref{fig.motivation.index}.

\SubParagraph{Composition-friendly.}
In order to support multi-column indexes covering multiple attributes (e.g., \emph{QTY} and an additional \emph{Date}), bitmap indexes create an instance for each attribute and apply bitwise ANDs/ORs among the corresponding bitvectors in both instances.
In contrast, tree-based indexes commonly build a single index for multiple attributes by using composite search keys (e.g., <\emph{QTY}, \emph{Date}>) \cite{Silberschatz2020}.
In this paper, we demonstrate that with large datasets, composition-based bitmap indexes are faster than tree-based indexes.
The reason is that bitmap indexes have a smaller memory footprint, and queries mainly perform sequential memory access to bitvectors, incurring fewer TLB and LLC misses (\S\ref{sec.eva.olap.dbx1000}).


\vspace{-0.06in}
\section{Why Bitmap Indexes Do Not Scale}\label{sec.challenge}
\vspace{-0.02in}

Prior updatable bitmap indexes are single-threaded and do not allow UDIs and queries to execute simultaneously.
We now discuss how to parallelize these designs for modern multi-cores and analyze their performance bottlenecks to motivate \sns's design decisions.

\begin{figure}[t]
    \centering
    \vspace{-0.1in}
    \setlength{\belowcaptionskip}{-10pt}
    \hspace*{-0.8em}
    \subfloat[Parallelized UpBit.] {
        \includegraphics[width=0.45\columnwidth]{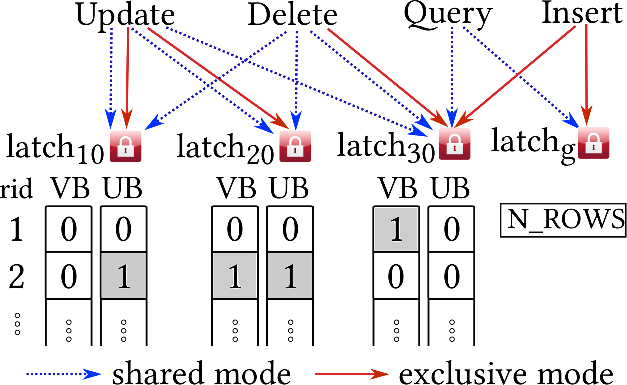}
        \label{fig.pain.concurrency}
    }
    \hspace*{-0.4em}
    \subfloat[Mean query and UDI latency.] {
        \includegraphics[width=0.5\columnwidth]{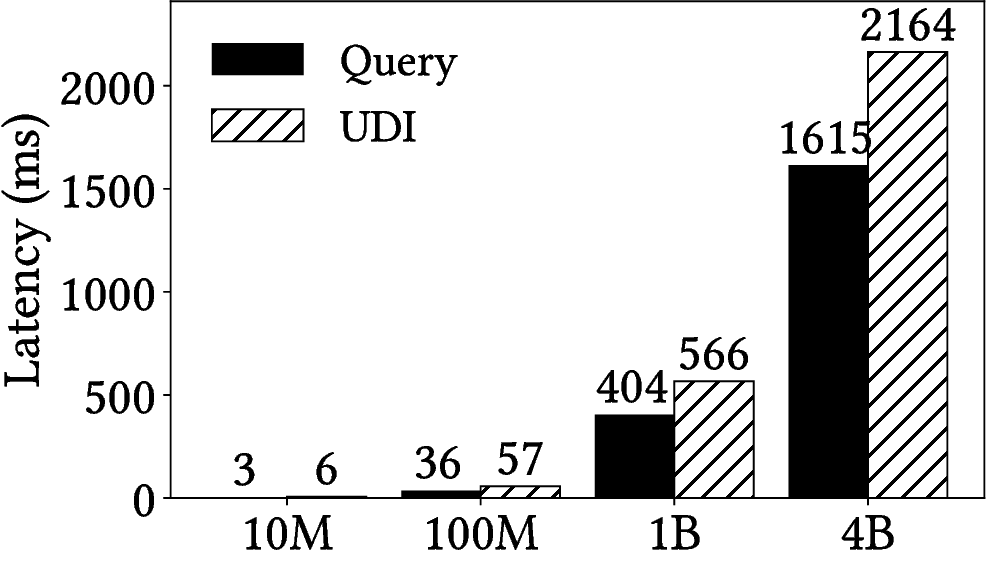}
        \label{fig.pain.bitvector}
    }
    \vspace{-0.06in}
    \caption{Existing updatable bitmap indexes do not scale because 
            (a) contention arises for high concurrency and 
            (b) update costs increase with large datasets.}
    \label{fig.pain}
    \vspace{-0.06in}
\end{figure}

\Paragraph{Bitmap Index Parallelization.}
\ub can be parallelized using a fine-grained synchronization mechanism, as illustrated in Figure \ref{fig.pain.concurrency}.
Specifically, the <VB, UB> pair of every value $v$ is protected by a reader-writer latch, denoted $latch_{v}$.
Global variables like \emph{N\_ROWS} are protected by a global latch $latch_{g}$.
Update and delete operations first acquire the $latch_{v}$ of all values in shared mode to retrieve the current value of the specified row. 
Then, they upgrade $latch_{v}$ of the corresponding bitvectors to exclusive mode in order to flip the necessary bits.
An insert acquires $latch_{g}$ and the corresponding $latch_{v}$ in exclusive mode.
Consequently, a query operation acquires $latch_{g}$ and the corresponding  \vn{$latch_{v}$} in shared mode.

Parallelizing other updatable bitmap indexes involves coarse-grained global latches, and thus, the parallelized algorithms suffer from higher contention.
We refer interested readers to Appendix.

\Paragraph{Challenges.}
Parallelized updatable bitmap indexes scale poorly and incur high tail latency (\S\ref{sec.eva.scale}) due to the following three challenges.
    
\SubParagraph{\textbf{(C1)} High Contention.} 
Queries and UDIs access the same bitvectors simultaneously, leading to contention (Figure~\ref{fig.pain.concurrency}).
Each \ub operation accesses two or more memory locations and thus acquires multiple latches in shared (blue dashed) and/or exclusive mode (red solid).
For high concurrency, this leads to long chains, where each operation waits for the preceding one to complete.

\SubParagraph{\textbf{(C2)} Long Critical Sections.} 
A query typically involves decoding and evaluating a bitvector, all under the protection of a latch.
As a result, latches on bitvectors are \textit{heavyweight} because their critical sections can last milliseconds or even seconds, which is orders of magnitude longer than typical critical sections (e.g., append to the tail of a list).
Therefore, with large datasets, each query takes up to seconds and incurs severe UDI delays, as shown in Figure \ref{fig.pain.bitvector}.
Further, UDIs can incur severe query delays because of the expensive \emph{decode-flip-encode} procedure.
 
\SubParagraph{\textbf{(C3)} Hot Bitvectors Are a Bottleneck.} In practice, the distribution of indexed attributes is not always uniform, such that a few bitvectors have much more 1s than the remaining ones.
As a result, those bitvectors (a) are less compressible and lead to a more expensive decode-flip-encode cycle, and (b) are more likely to be updated by UDIs, incurring even higher contention on the associated latches.

\section{\sn Design}\label{sec.arch.overview}
\vspace{-0.02in}


We now discuss \sns's building blocks and API.

\Paragraph{API.}
\sn complies with the standard specification for database indexes \cite{Silberschatz2020} and supports \vn{Query}, \vn{Update}, \vn{Delete}, and \vn{Insert} operations.
The (point or range) \vn{Query} operation returns the result as a bitvector or a row ID list.
\vn{Update} retrieves the current value of the specified row (by checking all bitvectors) and updates that row to the new value.
\vn{Insert} appends the new value to the tail of the bitmap index.
Further, an internal \vn{Merge} operation propagates logged changes to VBs.
\sns's operations are \emph{atomic}.

\Paragraph{Forward Progress.}
The \vn{Query} operation is wait-free \cite{Herlihy1991} with guaranteed completion, in contrast to state-of-the-art tree-based indexes where UDIs may interrupt overlapping range queries and force them to restart \cite{Levandoski2013}.
For UDIs, we implemented two versions:
the basic version, \sns-lk, that employs a latch to synchronize concurrent UDIs, and a latch-free version, \sns-lf, that adopts the \textit{helping mechanism}~\cite{Arbel-Raviv2018} and avoids UDIs blocking each other.

\vspace{-0.06in}
\subsection{Horizontal Update Delta (HUD)}\label{sec.overview.hud} 
\vspace{-0.02in}

Similar to the basic bitmap indexes (\S\ref{sec.bitmap.bitmap}), \sn associates each value of the indexed attribute to a compressed VB.
It avoids expensive decode-flip-encode cycles for each UDI by storing update information in an out-of-place row-wise manner and merging it into VBs lazily. 
Note that, by row-wise, we refer to a row of the conceptual bit-matrix of the bitmap index.
Our design, termed \emph{Horizontal Update Deltas (HUDs)}, enables efficient atomic UDIs without locking entire bitvectors, and hence addresses the high contention challenge in prior designs (\textbf{C1} from \S\ref{sec.challenge}).
HUD is key for high concurrency, by offering an effective organization of UDI deltas.

\Paragraph{Organization of HUD.}
Each HUD is conceptually a bit-array with a length equal to the cardinality of the domain.
In the general case, it has only 0s (i.e., the row has not been updated), and 1s accumulate with updates.
The number of 1s in a HUD is rarely more than two, as we discuss below.
Therefore, we compact each HUD as a list of positions, \textit{<row\_id, n\_ones, p1, p2, ...>}, where n\_ones is the number of 1s in the raw HUD, followed by their positions.

\begin{figure}[t]
    \centering
    \vspace{-0.1in}
    \setlength{\belowcaptionskip}{-10pt}
    \includegraphics[width=0.82\columnwidth]{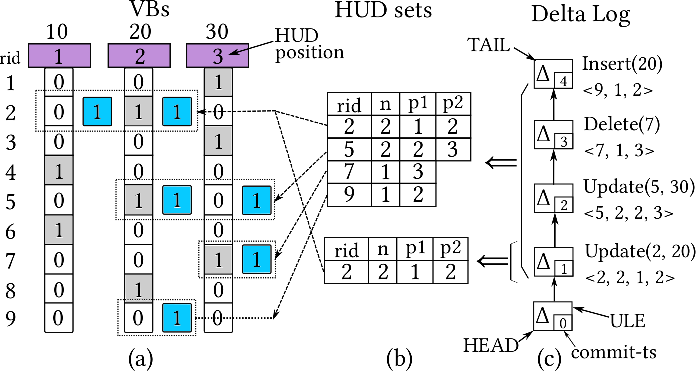}
    \vspace{-0.1in}
	\caption{
       \sn records UDIs in the form of HUDs that are organized chronologically in the per-index Delta Log (c). 
       Each query traverses a portion of the Delta Log and retrieves an on-demand HUD set (b).
       When applied to VBs, different HUD sets result in different snapshots of the index (a).}
	\label{fig.hub}
	\vspace{-0.1in}
\end{figure}

Figure \ref{fig.hub}a illustrates a bitmap index initially equivalent to the one shown in Figure~\ref{fig.bitmap}a.
An update operation of the $2^{nd}$ row to value 10 would flip the VBs containing the old and the new value.
Instead, \sn simply logs the delta information in the HUD <2, 2, 1, 2>.
Queries fetching this HUD apply it to the bitmap index by flipping the $2^{nd}$ bits of both VBs of values 10 and 20.
Similarly, updating the $5^{th}$ row to value 30, deleting the $7^{th}$ row, and inserting value 20 are executed by storing their delta information (Figure \ref{fig.hub}c), which will be applied to the queried VBs on demand.

\Paragraph{Benefits.}
Using HUDs brings the following benefits.
First, it enables lightweight atomic UDIs.
By gathering the information of UDIs performed on each row, which was scattered among different VBs, UDIs do not lock entire bitvectors.
Second, HUDs allow for efficient snapshotting for the underlying bit-matrix.
HUDs maintain complete information on the completed UDIs, such that different HUD sets efficiently generate different snapshots of the index (\S\ref{sec.overview.mvcc}).

\Paragraph{HUD Has Small Size.}
A HUD, in most cases, has up to two bits set, as shown in Figure \ref{fig.hub}b.
However, specific interleavings of updates and merges may result in longer HUDs, with very low probability.
For example, when $cardinality = 128$, the probability that a HUD contains 2, 3, or 7 1s is $\frac{1}{2^{7}}$, $\frac{1}{2^{14}}$, and $\frac{1}{2^{42}}$. More details can be found in Appendix.

\vspace{-0.06in}
\subsection{Delta Log}\label{sec.overview.deltalog} 
\vspace{-0.03in}

\Paragraph{Organization.}
\sn stores the changes from UDIs in a log, called \emph{Delta Log}, which consists of (linked) instances of \textit{UDI log entries}, named \emph{ULE} (Fig. \ref{fig.hub}c). 
Each \emph{ULE} entry contains the UDI's execution timestamp (\emph{commit-ts}, in the bottom-right corner of each \emph{ULE}).
The corresponding operations and the generated HUDs are listed on the right-hand side of the \emph{ULEs}.
A UDI completes by appending its \emph{ULE} at the tail of Delta Log.
Two global pointers, \emph{HEAD} and \emph{TAIL}, point to the first and last \emph{ULEs} in the list.
The \emph{commit-ts} values monotonically increase from HEAD to TAIL, allowing \sn to retrieve a HUD set by traversing the list in one pass.
We initially insert a dummy ULE with time 0, ensuring the Delta Log is never empty and simplifying subsequent read operations.
Appending HUDs to a log shortens UDIs' critical sections (\textbf{C2} from \S\ref{sec.challenge}).

\Paragraph{Example.}
As a concrete example, in Figure \ref{fig.hub}c, HEAD points to the dummy \emph{ULE} that links to the \emph{ULE} (with \emph{commit-ts = 1}) corresponding to the update operation that changes the second row from value 10 to 20.
The subsequent Update operation changes the value of the $5^{th}$ row to 30, by appending a \emph{ULE} with a HUD <5, 2, 2, 3>.
Then, a delete operation removes the $7^{th}$ row, and an insert operation appends value 20 at the tail of the index, by respectively appending their \textit{ULEs}.
By traversing \emph{ULEs} in different timestamp ranges, queries can retrieve different HUD sets (Figure \ref{fig.hub}b) and thus different snapshots of the index (\S\ref{sec.overview.mvcc}).

The exact implementation of Delta Log is orthogonal to \sns.
Since the \emph{commit-ts} values of \emph{ULEs} monotonically increase, any data structure that implements the \emph{list} abstract data type can be used.
In this paper, we chose a singly linked list for ease of presentation.

\vspace{-0.06in}
\subsection{Merging HUDs to VBs}\label{sec.overview.merge} 
\vspace{-0.03in}

In order to help queries and UDIs avoid traversing a long list of \emph{ULEs} to retrieve the necessary HUD sets, \sn maintains multiple versions of each VB, previously merged with HUDs.

\begin{figure}[t]
    \centering
    \vspace{-0.1in}
    \setlength{\belowcaptionskip}{-10pt}
    \includegraphics[width=0.88\columnwidth]{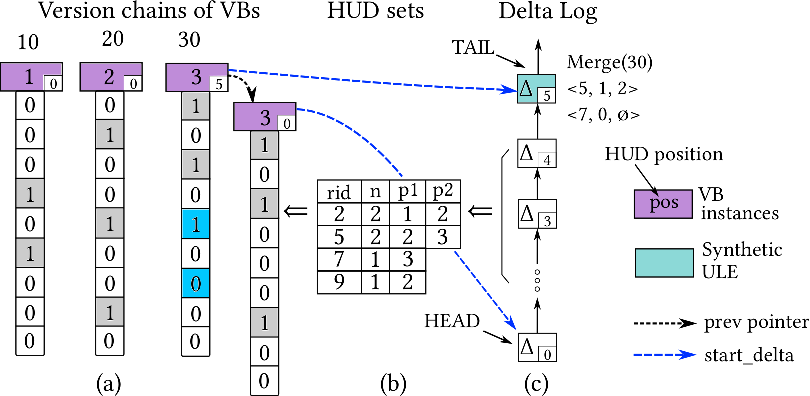}
    \vspace{-0.08in}
    \caption{A Merge operation generates a new VB, reducing the cost of applying deltas for subsequent queries and UDIs.}
    \label{fig.cubit-merge}
    \vspace{-0.1in}
\end{figure}

\Paragraph{Multi-Versioning VBs.} 
To this end, each value of the indexed domain is associated with a \emph{version chain}, i.e., a linked list of VB instances, as shown in Figure \ref{fig.cubit-merge}a.
Each VB instance contains its creation timestamp \emph{commit-ts} (bottom-right corner of each VB instance), a pointer \emph{start\_delta} to the ULE where subsequent queries and UDIs may start retrieving (not already merged) HUDs, and a pointer, \emph{prev}, linking to the previous version.
A query operation traverses the corresponding version chain (starting from the newest version), chooses the one with the largest \emph{commit-ts} that is less than or equal to the query's \emph{start-ts}, and then scans the Delta Log from the \emph{ULE} pointed to by the selected version's \emph{start\_delta}.

Figure \ref{fig.cubit-merge} illustrates a merge on value 30, starting from the index shown in Figure~\ref{fig.hub}.
\sn first makes a private copy of the latest version of the bitvector of value 30 (the version with \emph{commit-ts} $=$ 0 in Figure \ref{fig.cubit-merge}a), retrieves the HUD set shown in Figure \ref{fig.cubit-merge}b, and then merges the HUDs for value 30 into the private copy by flipping the $5^{th}$ and $7^{th}$ bits.
\sn then creates a new VB instance with an incremented \emph{commit-ts} and a \emph{prev} pointer referencing the old version of this VB. 
Its \emph{start\_delta} points to the newly-generated \emph{ULE} to skip the merged \emph{ULEs} and accelerate subsequent queries.
For example, queries with \emph{start-ts} > 5 on value 30 scan the Delta Log from \emph{ULE} with \emph{commit-ts} $=$ 5, rather than from \emph{HEAD}.

\Paragraph{Synthetic ULE.}
Upon merging a VB with a subset of HUDs from the Delta Log (flipping the bits indicated by HUDs associated with this VB), a new VB version is produced.
We now need to remove these specific entries from the Delta Log to ensure correctness.
However, in-place updating existing \emph{ULEs} would raise complicated synchronization issues. Instead, \sn appends a new \emph{synthetic ULE} that contains the HUDs that have not yet been merged, invalidating previous HUDs for those rows. 
For example, Figure~\ref{fig.cubit-merge}a shows the new version of VB for value 30 at timestamp 5. 
HUDs <5, 1, 2> and <7, 0, $\varnothing$> are generated and appended to the Delta Log, shown in Figure~\ref{fig.cubit-merge}c, indicating that the 1s that were in the $5^{th}$ and $7^{th}$ row have been merged into the newly-generated VB.
Note that the HUD <7, 0, $\varnothing$> can be omitted in practice.
Each synthetic \emph{ULE} corresponds to a successful merge operation, and each HUD in the synthetic \emph{ULE} corresponds to one merged bit. Note that ULEs are truncated when all their HUDs are invalidated.

\vspace{-0.06in}
\subsection{Segmented Bitvectors}\label{sec.overview.segbtv}
\vspace{-0.03in}

Unlike prior update-friendly bitmap indexes, \sn uses a layout where each bitvector is divided into \textit{multiple segments of fixed raw size} (bit count).
Each segment is then independently compressed using standard techniques like WAH. 

\Paragraph{Benefits.}
Even though segmentation has been widely used in prior research on both bitmaps~\cite{Lemire2018} and DBMSs~\cite{Dreseler2019, Kemper2011, Leis2014}, our mechanism is specifically designed for updatable bitmap indexes with the following unique features.
(1) It limits the cost of flipping bits of each UDI within one segment per bitvector (without any possible structure modification operations), hence reducing the cost of UDIs to a pre-defined threshold.
(2) By compressing corresponding segments using the same mechanism, it optimizes one of the most expensive operations: logical bitwise operations between segments. 

The combination of Delta Log (\S\ref{sec.overview.deltalog}), merging (\S\ref{sec.overview.merge}), and segmentation (\S\ref{sec.overview.segbtv}) keep the critical sections of updating bitvectors short, overall addressing challenge \textbf{C2} from \S\ref{sec.challenge}.

\vspace{-0.08in}
\subsection{Snapshotting}\label{sec.overview.mvcc} 
\vspace{-0.03in}

Internally, \sn utilizes multi-versioning with timestamp ordering to allow queries and UDIs to work on different snapshots, so that queries are guaranteed to complete even when UDIs are in progress. 
To this end, \sn maintains a global \emph{TIMESTAMP}, which is read at the beginning of every operation, denoted as \emph{start-ts}, and is incremented to reflect changes to the bitmap index (i.e., when UDIs complete or merge operations succeed).

\Paragraph{Multi-Versioning HUD sets.}
Given an operation's \emph{start-ts}, we can retrieve a HUD set by (1) traversing Delta Log from HEAD to the \emph{ULE} with the largest \emph{commit-ts} less than or equal to the operation's \emph{start-ts} and (2) collecting the HUDs in all of the \emph{ULEs}.
If HUDs for the same row appear in different \emph{ULEs}, the one committed lately is used to reflect the latest updates. 
Traversing the Delta Log can be significantly accelerated by merging HUDs into VBs (\S\ref{sec.overview.merge}) and moving HEAD forward. 

\Paragraph{Applying HUD Sets.}
Applying different HUD sets to the underlying bit-matrix results in different snapshots of the index.
Take Figure~\ref{fig.hub}b as an example. 
By traversing Delta Log, queries with \emph{start-ts = 1} use the bottom HUD set with 1 HUD, while queries with \emph{start-ts = 5} retrieve the top HUD set with 4 HUDs, capturing more UDIs.
When a HUD set is large, \sn first orders the HUDs according to their row ids, and then invokes multiple threads, each of which updates a subset of the bitvector's segments (\S\ref{sec.overview.segbtv}). 
By leveraging HUD sets, \sn offers lightweight snapshotting.

\vspace{-0.08in}
\subsection{Index Operations}\label{sec.overview.query} 
\vspace{-0.05in}

\Paragraph{Queries.} A query operation $Q$ on the value $val$ first retrieves a \emph{start-ts} by reading the global \emph{TIMESTAMP}, which essentially takes a snapshot of the bitmap index.
It then uses the \emph{start-ts} to search the version chain of the value $val$, and selects the bitvector with the largest \emph{commit-ts} that is less than or equal to \emph{start-ts}, which we refer to as the bitvector $V$.
For the delta information, it traverses the Delta Log and collects the HUDs in \emph{ULEs} with their \emph{commit-ts} $\in$ (\emph{V.commit-ts}, \emph{Q.start-ts}].
The search starts from the \emph{ULE} pointed to by the shortcut pointer \emph{Q.start\_delta}, which reduces the length of the traversal. 
The collected HUDs that do not affect value $val$ are filtered out.
If the result set is not empty, \sn makes a private copy of $V$, applys the delta information by flipping (in batch) the corresponding bits of the private copy, and then evaluates the resulting bitvector.
If the number of bits flipped is larger than a pre-defined threshold, the query operation sends a merge request to the background maintenance threads which attempt to reuse the newly-generated bitvector and insert it into the version chain.
If $V$ is segmented (\S\ref{sec.overview.segbtv}), \sn parallelized querying $V$ using the available CPU cores at a nearly linear speedup.

\Paragraph{UDIs.} Updates and deletes use \emph{start-ts} to retrieve the corresponding snapshot, similar to queries.
Using this snapshot, they read the old value of the specified row by checking each VB along with the associated HUD set.
This step is embarrassingly parallelized, by splitting the domain into ranges and assigning them to background threads.
Inserts omit this step.
A UDI logically flips the corresponding bits in the VBs.
Specifically, an update operation on the $n^{th}$ row would flip the $n^{th}$ bit of the VBs corresponding to the old and the new value, a delete operation would flip the $n^{th}$ bit of the deleted value, and an insert operation would set the last bit of the VB of the specified value $val$.
\sn actually generates a HUD for each UDI and appends it at the tail of Delta Log.

\vspace{-0.06in}
\subsection{Synchronization Mechanisms}\label{sec.overview.sync}
\vspace{-0.03in}

The lightweight snapshotting mechanism (\S\ref{sec.overview.mvcc}) allows queries (including range queries) to take snapshots of the whole index, such that \sns's queries are \emph{atomic} and \emph{wait-free}.
On the other hand, UDIs must be atomic with respect to other concurrent operations.
To this end, our basic algorithm employs a per-bitmap latch $latch_g$ to serialize the attempts to append \emph{ULEs} at the tail of Delta Log and increment TIMESTAMP.
Specifically, a UDI operation $U$ first attempts to grab $latch_g$.
Since $U$ works on a snapshot of the bitmap index, other operations (including merge operations) may have appended new \emph{ULEs} since $U$ started.
We thus check if there are any write-write conflicts between $U$'s HUDs and the set of HUDs in 
\emph{ULEs} with timestamps $\in$ (\emph{U.start-ts}, \emph{TAIL.commit-ts}].
Two HUDs conflict if they refer to the same row.
If there is a conflict, $U$ discards its HUD and restarts by reading with a new \emph{start-ts}.
Otherwise, $U$ appends its \emph{ULE} at the tail of Delta Log and increments TIMESTAMP by one.
It then releases the latch.
By incrementing the global TIMESTAMP, $U$ becomes visible to other operations.

\Paragraph{Overhead.}
Even though \sn employs $latch_g$, queries do not acquire it.
Moreover, the work in the critical section of this latch is orders of magnitude lighter compared to that of the latches employed by other bitmap indexes (thus, addressing \textbf{C2} from \S\ref{sec.challenge}). 
Our evaluation shows that \sn outperforms the alternatives in terms of throughput and (especially) tail latency (more details in \S\ref{sec.eva.scale}).

\Paragraph{Optimizations.}
Experimentally, we found that $latch_g$ can still raise high contention (see Appendix for an example with access skew).
We propose two mechanisms to alleviate the pressure from hot bitvectors and to address \textbf{C3} from \S\ref{sec.challenge}.

\SubParagraph{Consolidation Array.}
When Delta Log's contention is high, a \emph{consolidation array} approach \cite{Johnson2010, Moir2005, Shavit1997} is beneficial.
The basic idea is that when UDIs conflict, instead of busy-waiting, blocked UDIs consolidate their HUDs and delegate committing them in \emph{ULEs} to subsequent ones.
This allows a group of UDIs to consolidate into a single append operation, thus reducing contention.

\SubParagraph{Making \sns{} Latch-Free.}
We further introduce a \emph{helping} mechanism \cite{Michael1996, Arbel-Raviv2018} that makes \sns{} latch-free.
When UDIs and merges attempt to append \emph{ULEs} to the log simultaneously (i.e., write-write conflicts), they first \emph{help} the other complete and then retry, rather than competing with each other. 
Specifically, each UDI and merge operation records the old and new values of the variables to be updated in its \emph{ULE} before appending it to the tail of Delta Log using a \emph{CAS} instruction.
Once this step $O_1$ succeeds (i.e., this UDI operation linearizes), the \emph{ULE} becomes accessible to other threads via the next pointers of the \emph{ULEs} in Delta Log.
If another UDI or merge operation $O_2$ fails to append its \emph{ULE}, it first assists $O_1$ by retrieving the old and new values and updating the variable to the new value using \emph{CAS} instructions.
This approach allows \sns{} to prevent UDIs from blocking each other, addressing the primary cause of long tail latency in UDIs.
For implementation details and proof of correctness, we refer to Appendix.

\vspace{-0.08in}
\section{\sn implementation}\label{sec.arch.impl}
\vspace{-0.05in}

We now present the implementation details of \sn.

\Paragraph{ULE Pre-allocation.}
\sn pre-allocates an array of \emph{ULEs}, each of which is 32-byte long.
If a UDI wants a larger \emph{ULE}, which is very unlikely (\S\ref{sec.overview.hud}), it dynamically allocates a new \emph{ULE}; otherwise, it fetches a pre-allocated one.
This design benefits hardware prefetching because queries traverse Delta Log mostly sequentially.

\Paragraph{Timestamp Allocation.}
Allocating timestamps is frequently a performance bottleneck for multi-versioning \cite{Wu2017}.
\sn addresses this issue from two angles.
First, queries read but do not increment the global timestamp, dramatically reducing its contention.
Further, we use a hardware-based mechanism for timestamp allocation \cite{Kashyap2018} that provides a globally synchronized clock accessed as efficiently as reading hardware registers.
Overall, in our experimentation, timestamp allocation is not a performance bottleneck.

\Paragraph{Memory Reclamation.}
\sn's snapshotting provides higher concurrency levels, at the expense of increased memory footprint.
To address this, we proactively reclaim retired VBs that are no longer visible to worker threads and \emph{ULEs} once all of their HUDs have been merged into newly generated VBs.

Since VBs and \emph{ULEs} use commit timestamps (\textit{commit-ts}), it is natural for \sn to utilize an \emph{epoch-based} memory reclamation mechanism~\cite{RCU, Arbel-Raviv2018}.
Specifically, a version $A$ can be safely reclaimed, if (a) a newer version $B$ has been inserted at the head of its version chain, (b) the global TIMESTAMP has become equal to or larger than $B$'s commit-ts, and (c) each active worker thread has successfully performed at least one operation.
The correctness of this approach stems from the fact that the proposed epoch-based mechanism guarantees that before reclaiming $A$, there is at least one system-wide \emph{grace period} \cite{RCU}.
Therefore, subsequent operations will have \emph{start-ts} larger than or equal to $B$'s \textit{commit-ts}, and their queries and UDIs will not access $A$ anymore.

Similarly, a \emph{ULE} can be safely reclaimed, if (a) it is no longer accessible starting from any \emph{start\_delta} pointers of all VBs, and (b) each active worker thread has successfully performed at least one operation.
Once the head \emph{ULE} has been reclaimed, the global pointer HEAD is moved forward by the maintenance thread.

\Paragraph{Background Maintenance Threads.}
Memory reclamation is delegated to the background maintenance threads.
These threads are in charge of periodically (a) detecting invisible VBs and \emph{ULEs}, (b) detecting the grace period by utilizing a user-space implementation of RCU \cite{userspace-rcu} that helps \sn avoid explicitly tracking every active worker thread \cite{Triplett2011}, and (c) reclaiming retired objects.

\Paragraph{Operations between Bitvectors.}
DBMSs need to perform logical AND/OR operations between a group of \sn bitvectors.
We maintain the intermediate bitvector $IB$ as compressed when its bit density is below 0.2\% (highly compressible) to skip many 0-filled words in subsequent operations; otherwise, $IB$ is maintained as decompressed.
Meanwhile, when a bitvector to be merged has a bit density greater than 2\% (barely compressible), we decompress it before merging with $IB$.
This approach enables SIMD-based operations on 512-bit blocks, eliminating if-else branches and improving hardware prefetching.

\vspace{-0.06in}
\section{Experimental Evaluation}\label{sec.eva}
\vspace{-0.03in}

We demonstrate that \sn is fast, update-friendly, and scalable, and thus, it fits analytical queries on workloads with updates.


\Paragraph{Methodology.}
We experiment using In-place~\cite{Silberschatz2020}, UCB~\cite{Canahuate2007}, and \ubs~\cite{Athanassoulis2016a} as our baseline updatable bitmap indexes (\S\ref{sec.bitmap.bitmap}), and parallelize them using the most scalable strategies (\S\ref{sec.challenge}). 
We refer to the parallelized versions as In-place$^{+}$, UCB$^{+}$, and \ubs{}$^{+}$.
We assume that indexes are fully in-memory.

\Paragraph{Implementation.}
\sn and the baselines are implemented as C++ programs.
We compile the code with GCC 11.4 on Ubuntu 22.04, and use -O3 as our optimization level.
The open-source \emph{liburcu} \cite{userspace-rcu} is used as the framework of safe memory reclamation. 
Our artifacts implementing \sn and our changes to DBx1000 and DuckDB are available at \url{https://github.com/junchangwang/CUBIT}.

\Paragraph{Infrastructure.}
We experiment on a server with two Intel Xeon 5317 CPUs, each having 12 physical cores with Hyper-Threading running at 3.0GHz, and 18MB shared L3 cache. 
The system has 48 logical cores, 196GB DDR4 DRAM, and a 1TB SSD.


\Paragraph{Benchmarking Framework.}
We spawn a group of \emph{worker threads}, each of which executes the workload with the specified distribution of queries and UDIs.
Crossing NUMA does not noticeably affect (with an additional 3\% throughput overhead) bitmap indexes since they access bitvectors mostly sequentially.
We spawn up to 32 worker threads, each bound to a logical core, leaving the other 16 logical cores to the operating system and background maintenance threads.
Each experiment was repeated ten times, and the mean values were reported (standard deviation is less than 3\%).

\Paragraph{Workloads.}
We first use an in-house tool to generate integer data by varying three key properties:
data set size, domain cardinality, and data distribution (uniform or Zipfian).
We further test with industry-grade benchmarks, including the Berkeley Earth data \cite{BerkeleyEarthData2}, TPC-H \cite{TPCH}, and CH-benCHmark~\cite{Cole2011}.

\begin{figure}[t]
    \centering
    \vspace{-0.1in}
    \setlength{\belowcaptionskip}{-10pt}
    \hspace*{-0.8em}
    \subfloat[Overall throughput.] {
        \includegraphics[width=0.5\columnwidth]{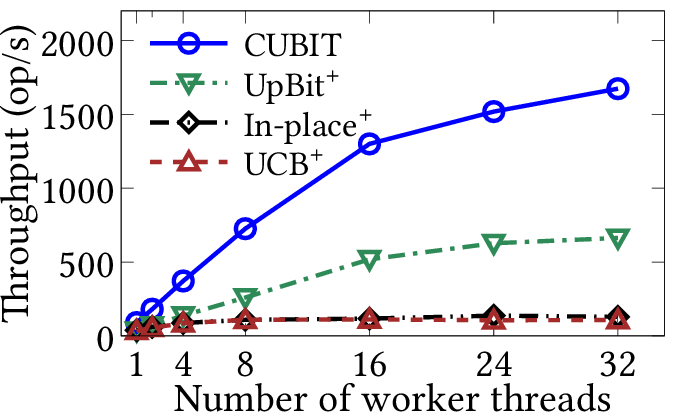}
        \label{eva.scale.thr}
    }
    \hspace*{-0.5em}
    \subfloat[Query latency.] {
        \includegraphics[width=0.5\columnwidth]{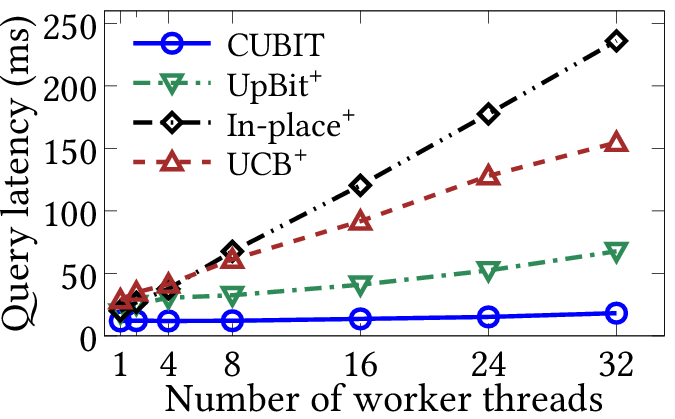}
        \label{eva.scale.query}
    }

    \hspace*{-0.8em}
    \subfloat[Update latency.] {
        \includegraphics[width=0.5\columnwidth]{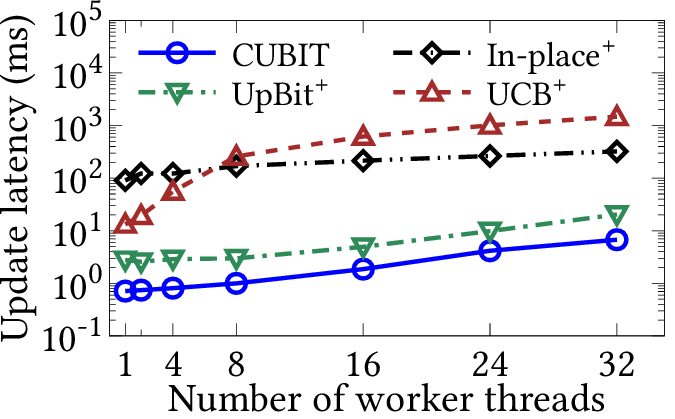}
        \label{eva.scale.update}
    }
    \hspace*{-0.5em}
    \subfloat[Insert latency.] {
        \includegraphics[width=0.5\columnwidth]{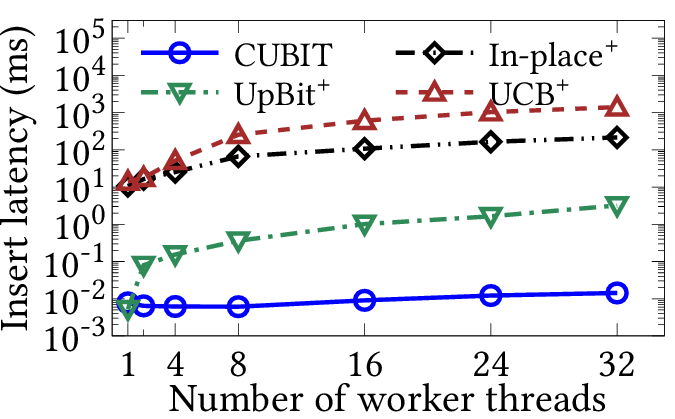}
        \label{eva.scale.insert}
    }

    \vspace{-0.1in}
    \caption{(a) \sn scales on multicore processors, and (b -- d) its query and UDI latency is almost constant for high concurrency. The y-axis of (c) and (d) are in log scale.}
    \label{eva.scale}   
    \vspace{-0.06in}
\end{figure}

\vspace{-0.06in}
\subsection{\sn Scales with Increased Parallelism} \label{sec.eva.scale} 
\vspace{-0.04in}

We evaluate the scalability of \sn and the baselines by varying the number of worker threads.
We simulate a typical use-case of updatable bitmap indexes \cite{Athanassoulis2016a}, where we have 90\% queries and 10\% UDIs, with a dataset of 100 million entries and domain cardinality equal to 100, and tune all approaches for this workload.
All bitmap indexes are tuned for this use-case.
For \sns, the merging threshold is 16, the number of segments per bitvector is 1000.
Figure \ref{eva.scale} compares the overall throughput, along with the response time of each type of operation. 
We make the following observations.

\Paragraph{\sn is Faster and Scales Better Than the Baselines.}
Figure \ref{eva.scale.thr} shows that In-place$^{+}$ and UCB$^{+}$, both of which are parallelized using global reader-writer latches, cannot scale with the number of worker threads.
By utilizing fine-grained, per-bitvector reader-writer latches, \ubs$^{+}$ scales better but plateaus for 16 or more threads.
This demonstrates that at high concurrency levels, the fine-grained locking mechanism in \ubs$^{+}$ becomes the bottleneck. 
The throughput of \sns, in sharp contrast, increases nearly linearly.
With 32 threads, \sn has 2.7$\times$ (13$\times$, 15.5$\times$) higher throughput than \ubs$^{+}$ (In-place$^{+}$, UCB$^{+}$).

\Paragraph{\sn Offers Fast and Scalable Read Performance.}
Figure~\ref{eva.scale.query} shows that \sns outperforms single-threaded \ub for queries because merging a HUD set into a UB in \sn is more lightweight than merging a <VB, UB> pair in \ub.
For high concurrency, the query latency of both In-place$^{+}$ and UCB$^{+}$ increases sharply because of the increased contention on the global reader-writer latch.
\ubs$^{+}$ distributes the contention to a group of fine-grained latches, improving query latency.
\sn, however, outperforms all alternatives. In particular, with 32 threads, the mean latency of \sn is 3.9$\times$  (8.5$\times$, 13.1$\times$) faster than \ub$^{+}$ (UCB$^{+}$, In-place$^{+}$). 

\Paragraph{\sns's UDIs are Fast.}
Figure~\ref{eva.scale.update} shows that for singled-threaded execution, \sns's updates outperform \ub because \sn simply appends the update information to its Delta Log, while \ub must decode-flip-encode the corresponding UBs.
Irrespective of the number of worker threads, each update and delete operation of \sn takes less than 10 ms (Figure~\ref{eva.scale.update}), and each insert takes about 0.01 ms (Figure~\ref{eva.scale.insert}).
Note that since updates and deletes follow the same general trends, Figure ~\ref{eva.scale.update} only shows results for updates.
The mean UDI latency of \sn is 3.0$\times$, 48.1$\times$, and 220.4$\times$ faster than \ubs$^{+}$, In-place$^{+}$, and UCB$^{+}$, respectively.

\vspace{-0.02in}
\subsection{Sensitivity Analysis} \label{sec.eva.factors} 
\vspace{-0.02in}

We now present a sensitivity analysis on domain cardinality, workload composition, and data size.
The experimental setup is the same as in \S\ref{sec.eva.scale},
with 16 worker threads and the best tuning for the baseline designs.

\begin{figure}[t]
    \vspace{-0.1in}
    \setlength{\belowcaptionskip}{-8pt}
    \hspace*{-0.8em}
    \subfloat{
        \includegraphics[width=0.5\columnwidth]{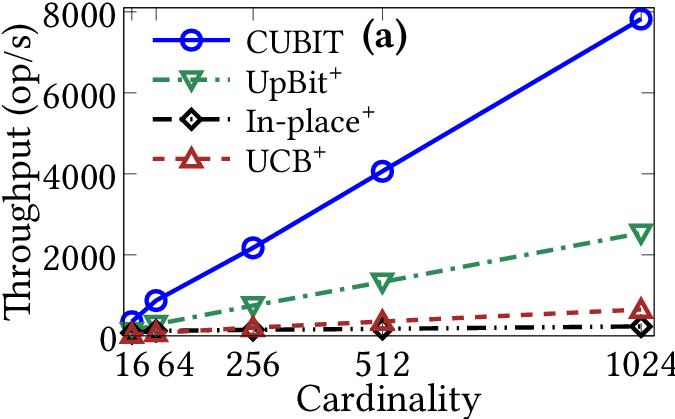}
        \label{eva.perf.cardinality.throughput}
    }
    \hspace*{-0.5em}
    \subfloat{
        \includegraphics[width=0.5\columnwidth]{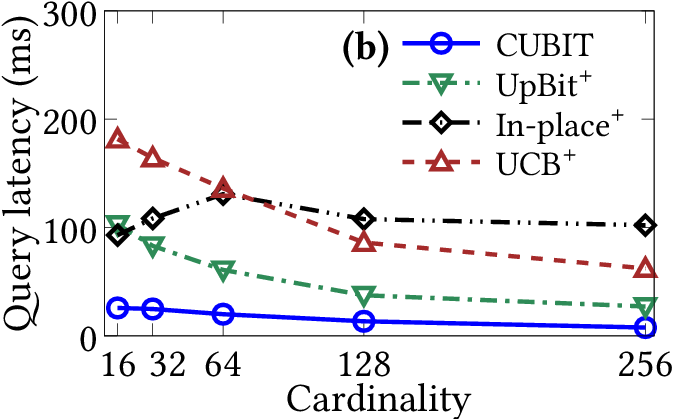}
        \label{eva.perf.cardinality.latency}
    }
    \vspace{-0.1in}
    \caption{Varying domain cardinality does not noticeably affect \sns's (a) throughput and (b) query latency.}
    \label{eva.perf.cardinality}      
    \vspace{-0.1in}
\end{figure}

\Paragraph{Impact of Domain Cardinality.}
Figure \ref{eva.perf.cardinality.throughput} shows that as the cardinality increases, the overall throughput of the tested approaches also increases.
The reason is that a larger cardinality leads to more compressible bitvectors, hence a smaller memory footprint.
Further, a larger cardinality eases the contention on each bitvector.
This also leads to lower query and UDI latency, as shown in Figure \ref{eva.perf.cardinality.latency}, which zooms to cardinality between 16 and 256.
The only outlier is In-place$^{+}$ because of the locking contention to append 0s and 1s.

\begin{figure}[t]
    \vspace{-0.1in}
    \setlength{\belowcaptionskip}{-4pt}
    \hspace*{-0.8em}
    \subfloat {
        \includegraphics[width=0.5\columnwidth]{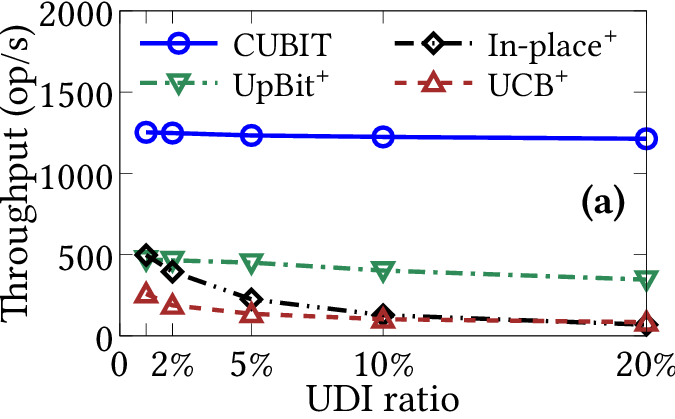}
        \label{eva.perf.udi.throughput}
    }
    \hspace*{-0.5em}
    \subfloat {
        \includegraphics[width=0.5\columnwidth]{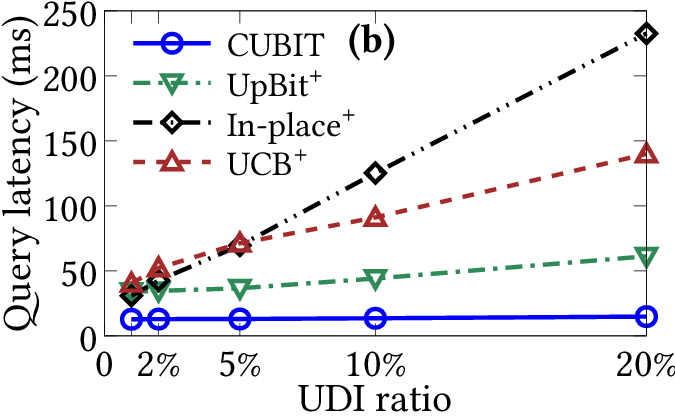}
        \label{eva.perf.udi.latency}
    }
    \vspace{-0.1in}
    \caption{(a) Throughput and (b) latency of \sn remain with variable UDI ratio.}
    \label{eva.perf.udi}      
    \vspace{-0.1in}
\end{figure}

\Paragraph{Impact of UDI Ratio.}
As the UDI ratio increases, the throughput of both In-place$^{+}$ and UCB$^{+}$ decreases sharply (Figure \ref{eva.perf.udi.throughput}).
The main reason is the increased contention on the global latches that increases query latency (Figure \ref{eva.perf.udi.latency}). 
\ubs's and \sn's throughput decreases as well. 
However, the performance loss is less than In-place$^{+}$ and UCB$^{+}$.
\ubs$^{+}$ faces increased contention on the reader-writer latches protecting bitvectors, while \sn faces increased contention on the latch protecting Delta Log.
Since the duration of the critical sections of \sn is orders of magnitude shorter than the baselines, \sn outperforms them in terms of both throughput and latency.
Note that the performance trends in Figure \ref{eva.perf.udi} continues until the UDI ratio becomes larger than 80\%.

\begin{figure}[t]
    \centering
    \vspace{-0.1in}
    \setlength{\belowcaptionskip}{-10pt}
    \hspace*{-0.8em}
    \subfloat {
        \includegraphics[width=0.5\columnwidth]{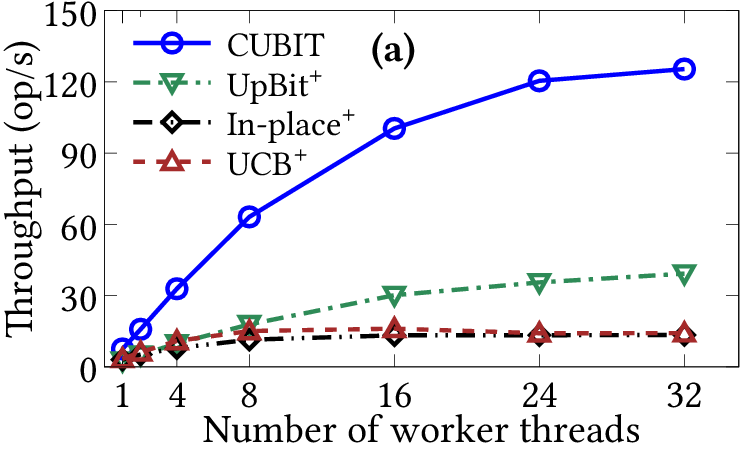}
        \label{eva.1B.throughput}
    }
    \hspace*{-0.5em}
    \subfloat {
        \includegraphics[width=0.5\columnwidth]{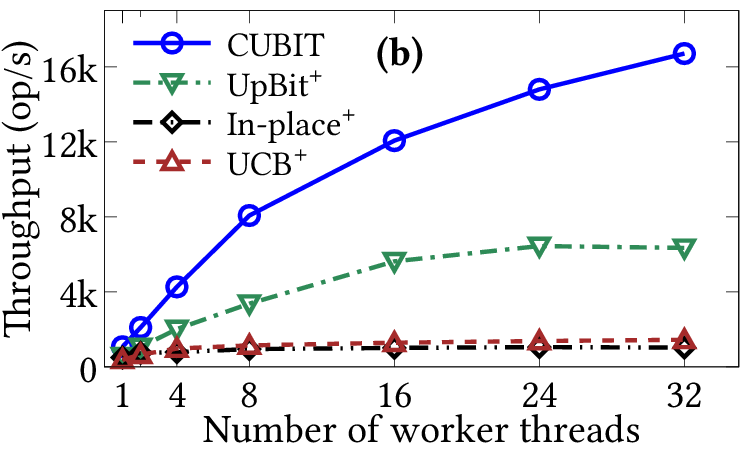}
        \label{eva.earth.throughput}
    }
    \vspace{-0.12in}
    \caption{(a) When increasing the dataset size (1B entries), and (b) when querying real datasets
            (Berkeley Earth dataset with 31M tuples and cardinality of 114),
            the relative behavior of all approaches remains the same.}
    \label{eva.largeData}
\end{figure}

\Paragraph{Impact of Data Size.} 
As the dataset size increases, the relative behavior of different indexes remains the same.
Figure \ref{eva.1B.throughput} shows the evaluation results with datasets containing 1B entries (cardinality $=$ 100). 
Figure \ref{eva.1B.throughput} looks almost identical to Figure \ref{eva.scale.thr}, which demonstrates that data size does not affect the performance trends and the relative behavior of different algorithms.
In this evaluation, however, each bitvector contains more bits.
Therefore, the absolute performance decreases nearly linearly as the dataset size increases.

\Paragraph{Berkeley Earth Dataset.}
For a real-life application, we evaluate \sn and its competitors using the Berkeley Earth dataset.
It is an open dataset for a climate study that contains measurements from 1.6 billion temperature reports, each of which contains information including temperature, time of measurement, and location.
From the Berkeley Earth data, we extract a dataset containing 31 million entries with cardinality 144.
Figure \ref{eva.earth.throughput} shows that with 32 worker threads, \sn's throughput is about 2.6$\times$, 11.5$\times$, and 16.2$\times$ 
higher than that of \ubs$^{+}$, UCB$^{+}$, and In-place$^{+}$, respectively.

\Paragraph{Impact of Data Skew.}
The distribution of data among bitvectors plays a key role in performance for two reasons.
First, biased distributions may lead to few \emph{target} bitvectors containing many more 1s than others, making them less compressible.
Second, the target bitvectors face higher contention levels among concurrent UDIs.
Our evaluation results show that for all bitmap indexes, UDI latency increases for skewed data because most of the UDIs involve a few bitvectors, leading to high contention on them.
However, \sn remains the most stable design because the \textit{helping mechanism} reduces the number of latches acquired, reducing \sns's P99 UDI latency (see Appendix for details).

\vspace{-0.02in}
\subsection{Tuning \sns}\label{sec.eva.tuning}
\vspace{-0.02in}

We now discuss the impact of \sns's tuning knobs and show that static tuning delivers robust performance.

\begin{figure}[t]
    \centering
    \vspace{-0.1in}
    \setlength{\belowcaptionskip}{-10pt}
    \hspace*{-0.8em}
    \subfloat{
        \includegraphics[width=0.5\columnwidth]{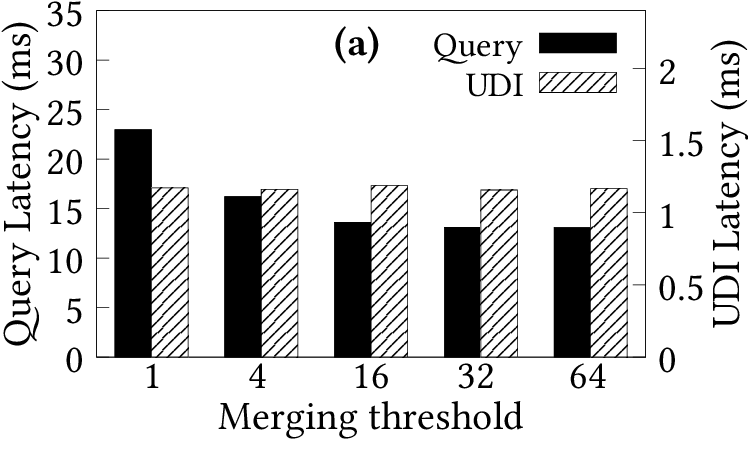}
        \label{eva.merge.10}
    }
    \hspace*{-0.6em}
    \subfloat{
        \includegraphics[width=0.5\columnwidth]{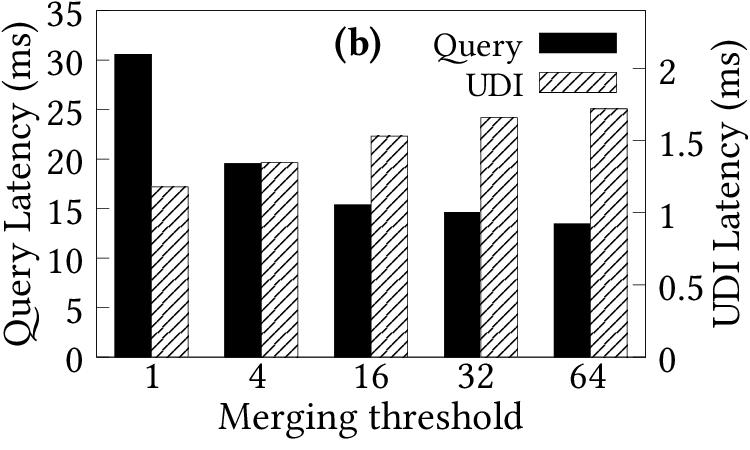}
        \label{eva.merge.20}
    }
    \vspace{-0.15in}
    \caption{For workloads with (a) 10\% UDIs and (b) 20\% UDIs,         
        \sns's query latency (left axis) decreases until the merging threshold becomes larger than 16,
        and UDI latency (right axis) slightly increases as a function of the merging threshold.}
    \label{eva.merge}      
    \vspace{-0.05in}
\end{figure}

\Paragraph{Merging HUDs into VBs.}
As UDIs accumulate, queries and UDIs need to traverse a long list in Delta Log.
\sn thus merges HUDs into VBs, as described in \S\ref{sec.overview.merge}.
Figures \ref{eva.merge.10} and \ref{eva.merge.20} show the average query and UDI latency as we vary the merging threshold, for the workloads with 10\% and 20\% UDIs, respectively.
In the general case, the query latency decreases as the merging threshold increases because of the reduced frequency of generating new VBs in query operations (\S\ref{sec.overview.query}).
However, when the threshold is larger than 16, the query latency plateaus because of the increased cost of traversing the list.
On the other hand, for both workloads, we observe a small increase in UDI latency as the merging threshold increases, because a UDI must traverse the Delta Log. 
This trend, however, is almost hidden because the update latency is dominated by the cost of retrieving the old value of the updated row.
Therefore, we set the merging threshold to 16 in our experiments.

\begin{figure}[t]
    \centering
    \vspace{-0.06in}
    \setlength{\belowcaptionskip}{-10pt}
    \hspace*{-0.8em}
    \subfloat{
        \includegraphics[width=0.5\columnwidth]{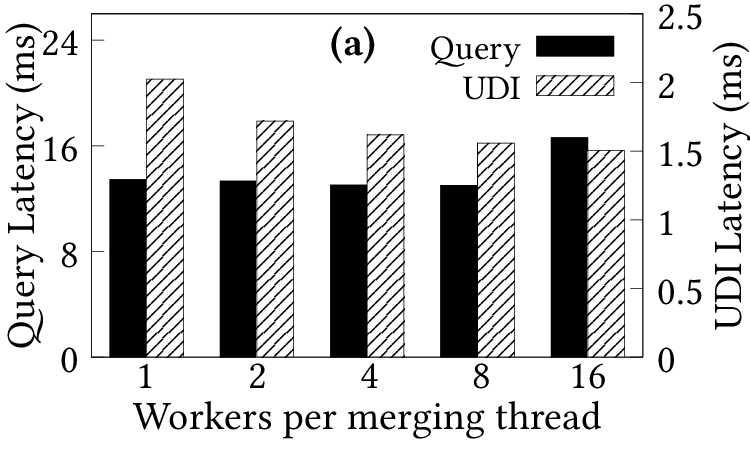}
        \label{eva.maint.10}
    }
    \hspace*{-0.5em}
    \subfloat{
        \includegraphics[width=0.5\columnwidth]{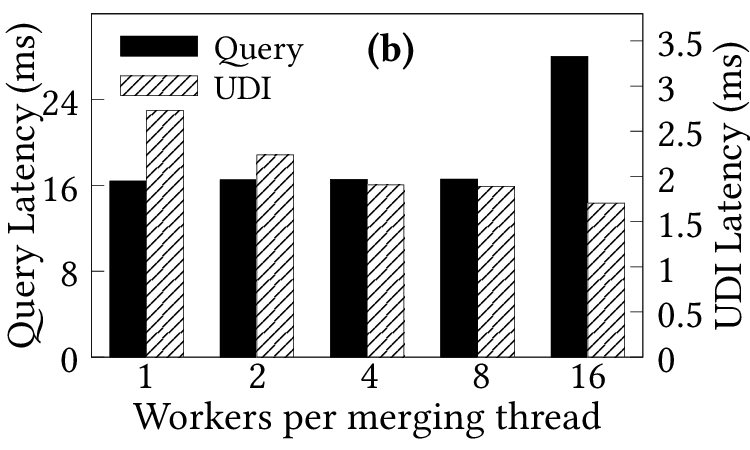}
        \label{eva.maint.20}
    }
    \vspace{-0.15in}
    \caption{For workloads with (a) 10\% UDIs and (b) 20\% UDIs,
        \sns's query latency plateaus until the ratio of worker and maintenance threads becomes larger than 8.
        UDIs always benefit from an increased ratio.}
    \label{eva.maint}      
    \vspace{-0.07in}
\end{figure}

\Paragraph{Number of Maintenance Threads.}
\sn offloads operations like garbage collection and merge operations to background maintenance threads.
Thus, it is critical to know in advance how many maintenance threads are enough.
Figure \ref{eva.maint} shows the average query latency and UDI latency as we vary the ratio of worker threads and maintenance threads, for workloads with 10\% (Figure~\ref{eva.maint.10}) and 20\% (Figure~\ref{eva.maint.20}) updates.
For both workloads, UDIs benefit from an increased ratio (fewer merging threads) because of a decreased contention on Delta Log.
The query latency plateaus when the ratio is less than eight.
However, when the ratio becomes larger than eight, query latency increases sharply because the maintenance threads become the bottleneck.
We thus set the ratio as 4 and create one maintenance thread for every four worker threads.

\begin{figure}[t]
    \centering
    \vspace{-0.06in}
    \setlength{\belowcaptionskip}{-10pt}
    \hspace*{-0.8em}
    \subfloat{
        \includegraphics[width=0.5\columnwidth]{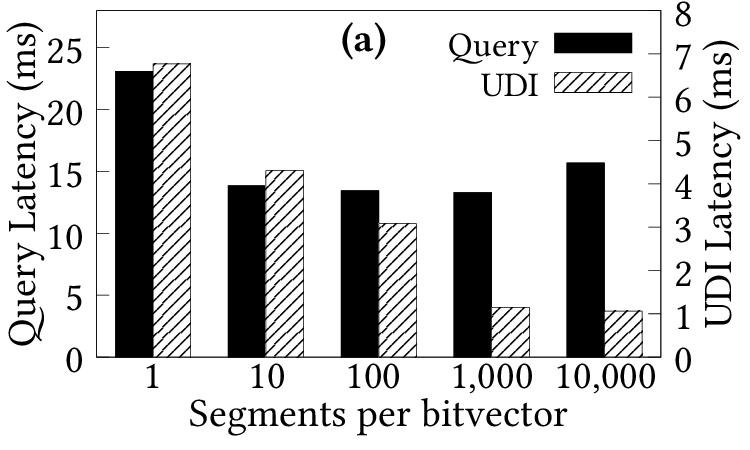}
        \label{eva.seg.10}
    }
    \hspace*{-0.5em}
    \subfloat{
        \includegraphics[width=0.5\columnwidth]{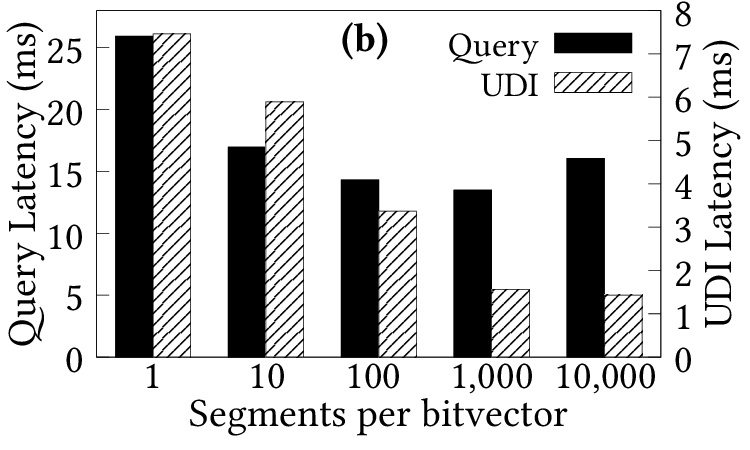}
        \label{eva.seg.20}
    }
    \vspace{-0.15in}
    \caption{For workloads with (a) 10\% UDIs and (b) 20\% UDIs,
        \sns's query latency decreases with finer segmentation granularity until each segment becomes less than 5KB.
        UDIs always benefit from a finer granularity.}
    \label{eva.seg}      
    \vspace{-0.1in}
\end{figure}

\Paragraph{Segments per Bitvector.}
\sn bitvector segmentation affects queries and UDIs in different ways.
Figure \ref{eva.seg} shows the average query and UDI latency for different numbers of segments of each bitvector (5MB in size), for workloads with 10\% (Figure~\ref{eva.seg.10}) and 20\% (Figure~\ref{eva.seg.20}) UDIs.
UDIs always benefit from a finer segmentation granularity that limits the execution of a UDI within a smaller segment.
However, the UDI latency plateaus when a bitvector is segmented into more than 1,000 segments (5KB-sized segments). 
On the other hand, a finer segmentation granularity allows a query to assign its tasks to a group of CPU cores (2 cores per query in this experiment) more evenly, significantly reducing query latency, as shown in Figure \ref{eva.seg}.
For more than 1,000 segments, query latency increases due to the cost of traversing the segments, thus, we use 1,000 segments in our experiments.

\begin{figure}[t]
    \centering
    \vspace{-0.2in}
    \setlength{\belowcaptionskip}{-10pt}
    \hspace*{-0.8em}
    \subfloat{
        \includegraphics[width=0.5\columnwidth]{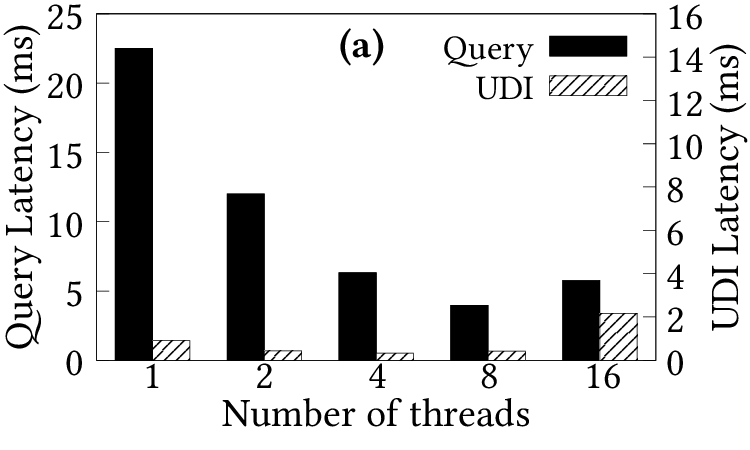}
        \label{eva.para.4}
    }
    \hspace*{-0.5em}
    \subfloat{
        \includegraphics[width=0.5\columnwidth]{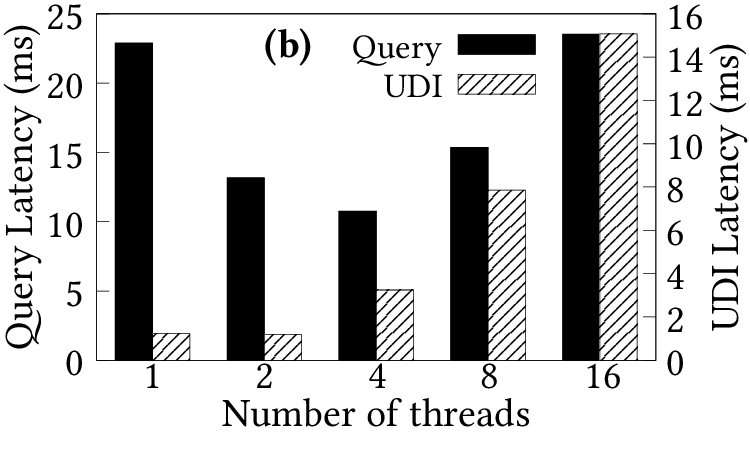}
        \label{eva.para.16}
    }
    \vspace{-0.2in}
    \caption{When there are (a) 4 and (b) 16 worker threads in the system,
        \sns's query and UDI latency decreases sharply as we assign more cores for each query, until the total required cores exceed the physical limit.}
    \label{eva.para}      
    \vspace{-0.1in}
\end{figure}

\Paragraph{Parallelism for Each Query.}
Using segmented bitvectors, queries can be embarrassingly parallelized.
This leads to a nearly linear speedup as long as the system has available CPU cores.
Figure \ref{eva.para} shows the average query and UDI latency as we vary the number of CPU cores for each query, for the experiments with 4 (Figure~\ref{eva.para.4}) and 16 (Figure~\ref{eva.para.16}) worker threads.
Figure \ref{eva.para.4} shows that using less than 8 cores per query, latency increases sharply.
However, for 16 cores (which means that \sn needs $4 \times 16 = 64$ cores in total), query and UDI latency increase sharply, as expected, since our server has 48 cores.
Similarly, Figure \ref{eva.para.16} demonstrates that for 16 worker threads, the number of cores used by each query should be less than 4 to not hurt query and UDI latency.
Based on these, we use two cores per query since our setup has 32 worker threads.

\vspace{-0.02in}
\subsection{\sn Benefits OLAP}\label{sec.eva.olap}
\vspace{-0.02in}

We now demonstrate that \sn benefits OLAP with batched updates by experimenting with the TPC-H benchmark with the standard refresh operations on DBx1000 (\S\ref{sec.eva.olap.dbx1000}) and DuckDB (\S\ref{sec.eva.olap.duckdb}).

\Paragraph{TPC-H.}
We experiment with TPC-H with scale factor (SF) = 10 and refresh operations (RF1 and RF2).
RF1 (RF2) inserts (deletes) 4,500 tuples in \emph{LINEITEM} in batch, and then updates related indexes.
The workload consists of 98\% Q6, 1\% RF1, and 1\% RF2.

\subsubsection{\textbf{DBx1000 Integration}}\label{sec.eva.olap.dbx1000}
In order to compare \sn with the alternative indexes at different concurrency levels, we integrate \sn into DBx1000~\cite{Yu2014}, a row-based prototype DBMS.

\Paragraph{Methodology.}
We implement Bw-Tree and ART based on their open and comparable implementations \cite{Wang2018a} and optimize their performance with our testbed.
In particular, our ART uses optimized arrays, rather than linked lists, to handle duplicate keys.
Overall, we compare:
(1) a parallelized \textbf{Scan},
(2) a \textbf{Hash} index with bucket size $B = 64K$,
(3) a \textbf{B$^+$-tree} index with a fixed fanout of 32,
(4) \textbf{Bw-Tree} with a maximum inner node size of 64 \cite{Levandoski2013} ,
(5) \textbf{ART} with a maximum node fanout of 256 \cite{Leis2013}, and
(6) \textbf{\sns}.
Since Q6 selects on \emph{l\_shipdate}, \emph{l\_discount}, and \emph{l\_quantity}, we built three \sn indexes.
Other indexes are built using multi-column indexes over the three attributes (the \textit{best-case} for the baselines).
The composite search keys (\emph{l\_shipdate}, \emph{l\_discount}, and \emph{l\_quantity}) for these indexes are not unique, so for SF $=$ 10, each index node contains, on average, 50 entries with the same key in the form of an array.

The Hash index is protected by fine-grained per-bucket reader-writer latches,
and the \bt and ART employ optimistic lock coupling \cite{Leis2016}.
Both Bw-Tree and \sn offer latch-free queries.
Apart from the Scan, each query first retrieves a set of tuple IDs through the indexes, and then fetches these tuples to calculate the query result.
Tree-based indexing could ensure a better access locality by sorting the IDs first.
This approach, however, requires CPU and memory resources.
Similarly to prior work~\cite{Kester2017}, we observed that sorting slows down tree-based indexing by $\sim$6\%.

\begin{figure}[t]
    \centering
    \setlength{\belowcaptionskip}{-10pt}
    \hspace*{-0.8em}
    \subfloat {
        \includegraphics[width=0.5\columnwidth]{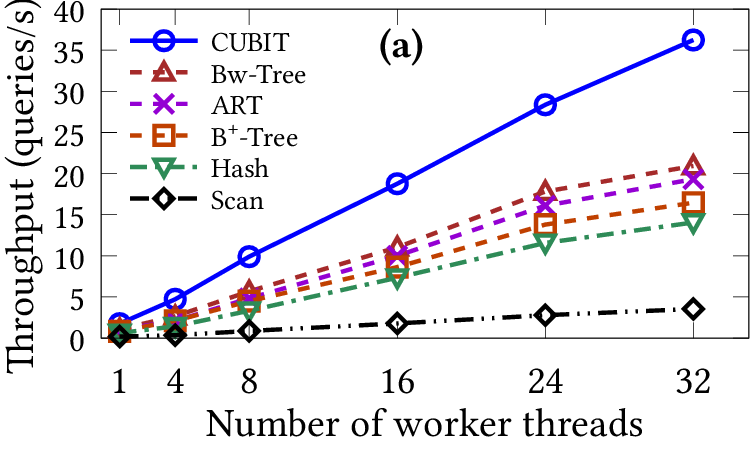}
        \label{eva.DBx1000.throughput}
    }
    \hspace*{-0.5em}
    \subfloat {
        \includegraphics[width=0.5\columnwidth]{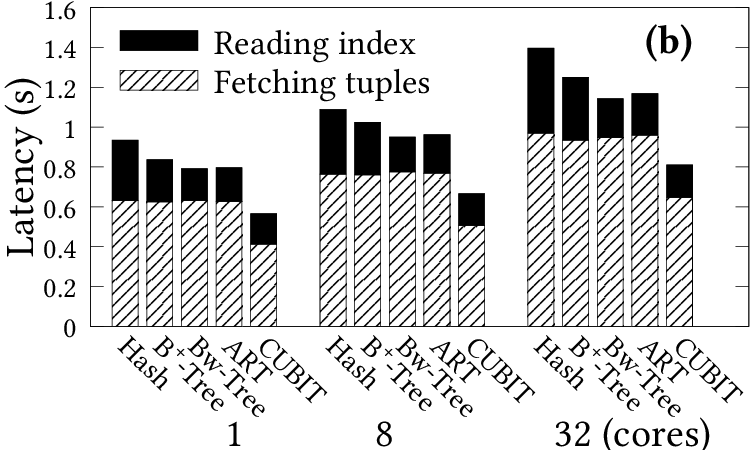}
        \label{eva.DBx1000.query.breakdown}
    }
    \vspace{-0.15in}
    \caption{(a) At different concurrency levels,
            DBx1000 with \sn outperforms all baselines for TPC-H Q6.
            (b) \sn index access cost is always smaller.
            Moreover, \sn provides an ordered ID list,
            leading to faster tuple fetches.}
    \label{eva.DBx1000}      
    \vspace{-0.08in}
\end{figure}

\Paragraph{Benefits.}
We observed the following benefits of using \sn.

\SubParagraph{(a) High Availability.}
\sn enables bitmap indexing for OLAP with batched updates, by overcoming the maintenance downtime \cite{Stockinger2009, Burleson2011}.
Our evaluation results show that RF1 and RF2 do not noticeably affect concurrent queries.

\SubParagraph{(b) Low Memory Footprint.}
The hash index, ART, \bts, and Bw-Tree, respectively, use 1.92GB, 1.93GB, 1.96GB, and 1.97GB of memory, mainly because they must maintain keys and pointers to the tuples, which is about 1.44GB.
In contrast, \sn size is only 156MB.

\SubParagraph{(c) High Performance.}
Despite the batched updates, \sn enables faster execution for selective queries, and the throughput of DBx1000 with \sn scales linearly with the number of worker threads, as shown in Figure \ref{eva.DBx1000.throughput}.
The scalability of DBx1000 with the baseline indexes suffers from random accesses and cache misses because the indexes are not clustered, while by construction, \sn always generates clustered base data accesses.
In particular, using 32 threads in DBx1000, \sn provides 1.7$\times$, 2.2$\times$, 2.2$\times$, 2.6$\times$, and 10.3$\times$ higher throughput than Bw-Tree, ART, \bts, Hash index, and Scan, respectively.
Scanning is the slowest approach because of its excessive data access and its high cache miss rate.

\begin{table}[bh]
    \centering
    \vspace{-0.1in}
    \captionsetup{font=small}
    \small
    \caption{Normalized last level cache misses and TLB misses of the two stages of each Q6.
         DBx1000 with \sn exhibits better spatial locality and experiences fewer misses.}
    \vspace{-0.1in}
    \begin{tabular}{|c|c|c|c|c|c|c|}
        \hline
        \multicolumn{2}{|c|}{\diagbox{Stages}{Indexes}}  & $\frac{\text{Bw-Tree}}{\text{\sns}}$ & $\frac{\text{ART}}{\text{\sns}}$ & $\frac{\text{B$^+$-tree}}{\text{\sns}}$ & $\frac{\text{Hash}}{\text{\sns}}$ & $\frac{\text{Scan}}{\text{\sns}}$ \\
        \hline
        \multirow{2}{*}{Reading index} & LLC & 1.4 & 1.3 & 1.7 & 1.9 & --- \\
        \cline{2-7}
        & TLB & 1.4 & 1.5 & 2.3 & 2.2 & --- \\
        \hline
        \multirow{2}{*}{Fetching tuples} & LLC & 5.4 & 5.5 & 6.3 & 5.8 & 342.7 \\
        \cline{2-7}
        & TLB & 4.0 & 3.9 & 4.3 & 4.4 & 12.2 \\
        \hline
    \end{tabular}
    \label{table.eva.perf}
    \vspace{-0.15in}
\end{table}

We inspect each indexing strategy by measuring the cost of the two stages of each Q6 -- probing the index and fetching tuples.
Figure \ref{eva.DBx1000}b demonstrates that the benefits of using \sn are two-fold.
On the one hand, probing \sn is faster for all concurrency levels.
For example, for the serial execution (\# worker thread $=$ 1), it takes \sn 152ms to retrieve the tuple ID list, which is 1.1 -- 3.0$\times$ faster than the alternative indexes.
The main reason is that \sn has a smaller memory footprint, and a query fundamentally performs sequential memory access to each bitvector, exhibiting better spatial locality.
This is demonstrated in Table \ref{table.eva.perf} in which we list the normalized last level cache (LLC)  and TLB misses of different stages of each Q6.
The first two rows show that reading Bw-Tree incurs 1.4$\times$ and 1.4$\times$ more LLC and TLB misses than reading \sns.
On the other hand, fetching tuples according to the ID list generated by \sn is faster.
For example, for the serial execution version, it takes DBx1000 413ms to fetch the tuples by using the ID list provided by \sn, 1.5$\times$ faster than by using the ID lists generated by the alternative indexes, including Bw-Tree and ART.
The main reason is that the tuple IDs provided by \sn are inherently ordered, while this is not generally the case for other indexes.
Essentially, by virtue of its construction, \sn is a clustered secondary index since the retrieved IDs follow the same order as the tuples in the base data. 
In contrast, accessing data with the hash and tree-based indexes leads to more cache and TLB misses.
This is shown in Table \ref{table.eva.perf} (last two rows): using the ID list generated by Bw-Tree, the DBMS incurs 5.4$\times$ and 4.0$\times$ more LLC misses and TLB misses than \sns.
Finally, Table \ref{table.eva.perf} shows that Scan performs poorly in row-based DBMSs because it accesses more data and incurs more cache misses.

\vspace{-0.05in}
\subsubsection{\textbf{DuckDB Integration}}\label{sec.eva.olap.duckdb}
To demonstrate the usefulness of \sns, we integrate \sn into DuckDB \cite{Raasveldt2019}, a column-based OLAP DBMS, that has been heavily optimized for analytical applications.
We optimized as many choke points in DuckDB's query engine as possible \cite{Boncz2013, Dreseler2020}.
For example, by excluding certain group-by attributes that can be derived from the primary key, we achieved a 46\% performance improvement in Q10.
We optimize its Scan operator by (1) storing column values more compactly (e.g., by compressing \emph{l\_quantity} from 13 to 6 bits) with a storage layout inspired by BitWeaving's HBP \cite{Li2013}, and (2) scanning columns using SIMD instructions, leading to a 36\% lower latency for Q6.
We refer to our optimized version as DuckDB$^+$ and use it as the baseline.

\sns's update-friendly design allows us to maintain \sn instances for frequently updated attributes
instead of repeatedly destroying and re-creating them \cite{Oracle2007}.
This not only reduces the amount of data read from storage for the \emph{Scan} operator, but also provides sufficient information for the \emph{Aggregation} and \emph{Join} operators, eliminating the need to build intermediate data structures (e.g., hash table for Joins).
We demonstrate this by using \sn to accelerate the \emph{Scan}, \emph{Aggregation}, and \emph{Join} operators (the top three choke points in DBMS query engines \cite{Boncz2013, Dreseler2020}) in DuckDB$^+$ as follows
(see Appendix for more details).

\SubParagraph{Scan.}
By maintaining \sn instances on the attributes involved in the WHERE clauses, our \sns-powered Scan operator reads fewer columns.
For example, in Q6, DuckDB$^+$ sequentially scans four columns (\emph{l\_extendedprice}, \emph{l\_shipdate}, \emph{l\_discount}, and \emph{l\_quantity}), which accounts for 95\% of the execution time.
In contrast, our operator generates a resulting bitvector, allowing it to probe only two columns, effectively halving the data read from storage.

\begin{figure}[t]
    \centering
    \vspace{-0.1in}
    \setlength{\belowcaptionskip}{-10pt}
    \hspace*{-0.8em}
    \includegraphics[width=\columnwidth]{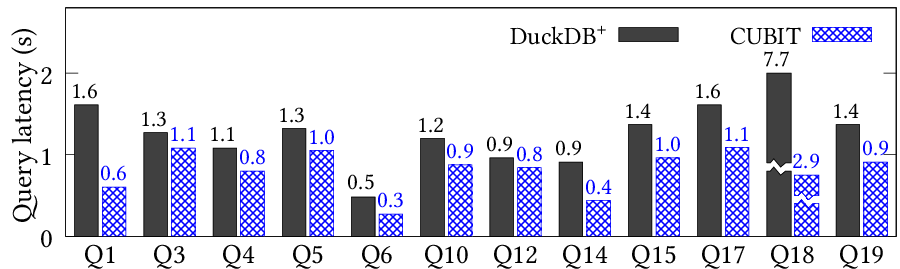}
    \vspace{-0.16in}
    \caption{Being update-friendly, \sn can be built on the LINEITEM and ORDERS tables that are updated by refresh transactions.
    By replacing the column-based scan, perfect-hash-based aggregation, and hash-based join operators of DuckDB$^+$, \sns-powered query engine outperforms for all of the queries.}
    \label{fig.eva.tpch}
    \vspace{-0.1in}
\end{figure}

\SubParagraph{Aggregation.}
By maintaining \sn instances on the attributes used as group-by factors (e.g., \emph{l\_returnflag} and \emph{l\_linestatus} for Q1), the \sns-powered Aggregation operator avoids reading these columns from storage.
Additionally, \sn eliminates the need to maintain data structures for aggregations by (1) determining the positions of matching tuples for each group-by category by \emph{AND}ing bitvectors from \sn instances, and (2) calculating the aggregations for each category by reading the specified entries in one pass---a computation mode amenable to SIMD instructions.

\SubParagraph{Join.}
By maintaining a \sn instance on the join attribute (e.g., \emph{l\_orderkey} for Q5), the \sns-powered Join operator avoids scanning one column data in the fact table.
Additionally, the \sn instance provides sufficient information to join two tables by using its bitvectors.
For example, for the join in Q5, our operator retrieves the \emph{orderkey} set from the ORDERS table, reads the corresponding bitvectors from the \sn instance, and performs logical ORs between them.
Using the resulting bitvector, the operator probes the fact table and performs aggregations in one pass.

With our optimized operators, we experiment with 12 out of 22 TPC-H queries, including scan-intensive (Q6 and Q12), aggregation-heavy (Q1 and Q18), and join-dominant (Q3, Q4, Q5, Q10, Q14, Q15, Q17, and Q19) queries.
We omit other queries that neither involve the fact table LINEITEM (e.g., Q2) nor present obvious choke points for query engines (e.g., Q16).
Our evaluation results show that \sns-powered query engine is 1.2--2.7$\times$ faster than the native approaches in DuckDB$^+$, shown in Figure \ref{fig.eva.tpch}.

\begin{figure}[t]
    \centering
    \vspace{-0.2in}
    \setlength{\belowcaptionskip}{-10pt}
    \hspace*{-0.8em}
    \subfloat {
        \includegraphics[width=0.48\columnwidth]{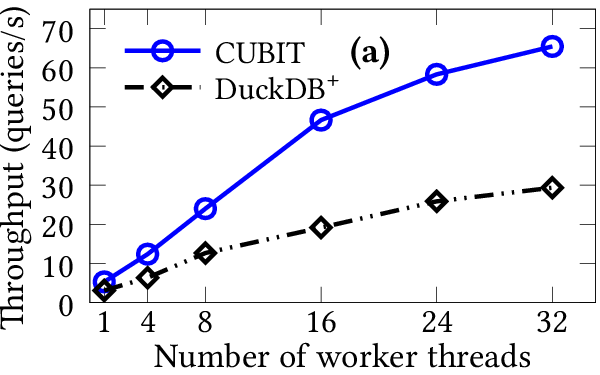}
        \label{eva.DuckDB.throughput}
    }
    \hspace*{-0.3em}
    \subfloat {
        \includegraphics[width=0.48\columnwidth]{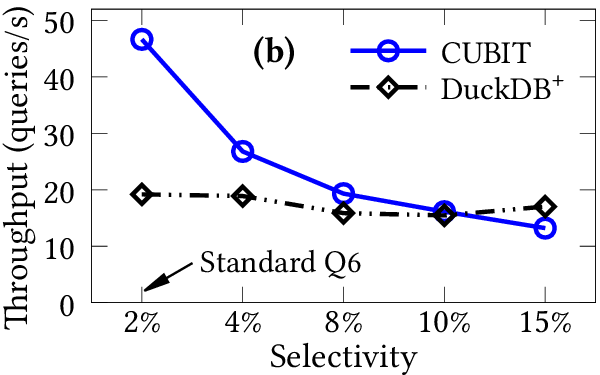}
        \label{eva.DuckDB.selectivity}
    }
    \vspace{-0.1in}
    \caption{(a) At different concurrency levels,
            \sns-based probe outperforms the highly-optimized scan in DuckDB$^+$ for TPC-H Q6.
            (b) This trend continues until the selectivity reaches 10\%.}
    \label{eva.DuckDB}      
    \vspace{-0.1in}
\end{figure}

We further explore the conditions under which maintaining \sn instances outperforms traditional column scanning by using Q6 as an example.
In Q6, we create three \sn instances on the attributes \emph{l\_shipdate} (cardinality = 2,526), \emph{l\_discount} (cardinality = 11), and \emph{l\_quantity} (cardinality = 50).
Our operator selects bitvectors for 365 days, 3 discounts, and 25 quantities, and performs ORs/ANDs between these 393 (365+3+25) bitvectors to compute the resulting bitvector (selectivity = 2\%), which is maintained in segments (\S\ref{sec.overview.segbtv}).
We use an AVX-512-based mechanism \cite{BitsetDecoding} to convert the resulting bitvector to an ID list, which is used to probe the \emph{l\_extendedprice} and \emph{l\_discount} columns.
We made the following observations:

\SubParagraph{\sn Scales for High Concurrency.}
Each bitvector segment covers a small range of physical pages, such that worker threads run in parallel, without any synchronization overhead, leading to near-ideal scalability (Figure~\ref{eva.DuckDB}a).

\SubParagraph{\sn Has Wide Applicability.}
The effectiveness of \sn makes it a potential replacement for Scan for several of use cases.
To demonstrate this, we use 16 cores but artificially increase the selectivity of Q6 by varying the combination of \emph{l\_shipdate} and \emph{l\_discount}.
\emph{\sn outperforms Scans for up to 10\% selectivity} (Figure \ref{eva.DuckDB}b).

\vspace{-0.03in}
\subsection{\sn Benefits HTAP}\label{sec.eva.htap}
\vspace{-0.04in}


\Paragraph{CH-benCHmark.}
We use DBx1000 because of its scalability and 
implement the CH-benCHmark~\cite{Cole2011} that consists of a full version of the TPC-C benchmark and a set of TPC-H-equivalent analytical queries on the same tables.
$T$ threads run TPC-C transactions that heavily update tables, including the \emph{ORDER-LINE} table, and the remaining $A$ threads run CH-benCHmark analytical queries on these tables.
We use an initial dataset of 100 warehouses, so \emph{ORDER-LINE} contains about 2.5GB of data and 30M tuples.
The TPC-C transactions insert $\sim$10M new tuples during each trial.

\Paragraph{Indexes.}
We built one \sn index for each attribute in the analytical queries (i.e., Q1 and Q6), and other indexes are built using multi-column indexes for each query, as in \S\ref{sec.eva.olap.dbx1000}.

\Paragraph{Concurrency Control.}
\sn can be used as a general secondary index when the DBMS has its concurrency control mechanism (e.g., 2PL).
An independent lock manager is responsible for resolving conflicting data updates \cite{Levandoski2011, Levandoski2013} and, thus, we focus on the atomicity of \sns's operations.
We also implement an MVCC mechanism, called \emph{Post-Timestamping MVCC (PTMVCC)}, that is similar to Hekaton~\cite{Diaconu2013} but only increments its TIMESTAMP when transactions with updates successfully commit~\cite{Kim2019}.
A transaction traverses the chains of tuples or speculatively reads other transactions' workspace to fetch the tuple versions visible to it.
PTMVCC reduces the contention on updating TIMESTAMP~\cite{Yu2014} and provides a stronger progress guarantee for analytical queries that are never blocked nor aborted~\cite{Kim2019}.
PTMVCC provides snapshot isolation.
When using 16 transactional and 4 analytical threads, PTMVCC achieves an 8\% performance improvement over the 2PL+\sn solution, primarily due to (1) the wait-free execution of queries even during ongoing updates, and (2) the elimination of tuple latching overhead.
We thus use PTMVCC in our evaluation.

\Paragraph{Selectivity.}
In our evaluation, we found that many attributes in CH-benCHmark cover a narrow scope, such that most queries select almost all tuples.
We thus modified the propagated values and the query predicates to provide variable selectivity.
For example, we set the values of the \emph{ol\_delivery\_d} attribute in the \emph{ORDER-LINE} table in the range of [1983, 2023), and the values of the \emph{ol\_quantity} attribute in the range of [1, 25000), both in a uniform distribution.
As a consequence, each CH-benCHmark Q1 selects rows on years (16 out of 40) and delivery state (9 out of 10), leading to an average selectivity of $\frac{16}{40}\times\frac{9}{10} \approx 36\%$, 
and each Q6 selects rows on years (20 out of 40), quantities (1000 out of 25,000), and delivery state (9 out of 10), leading to an average selectivity of $\frac{20}{40}\times\frac{1}{25}\times\frac{9}{10} \approx 1.8\%$.

\begin{figure}[t]
    \centering
    \vspace{-0.1in}
    \setlength{\belowcaptionskip}{-10pt}
    \hspace*{-0.8em}
    \includegraphics[width=\columnwidth]{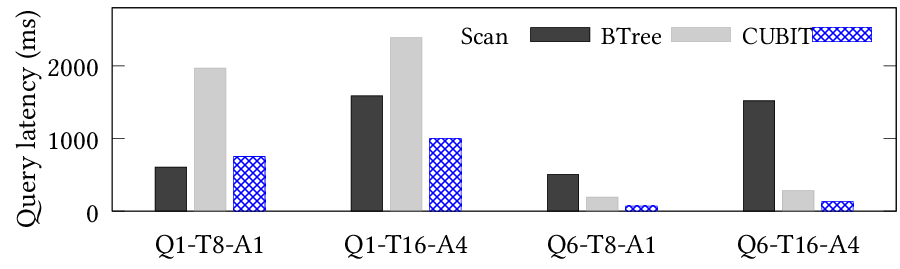}
    \vspace{-0.16in}
    \caption{For HTAP DBMSs, \sn can speedup analytical queries with high (Q1) and low (Q6) selectivity, irrespective of concurrent updates.
            Higher contention (16 TPC-C threads instead of 8) and parallelism (4 analytical queries instead of 1) do not noticeably affect \sns's performance.}
    \label{fig.eva.chbenchmark}      
    \vspace{-0.1in}
\end{figure}

\Paragraph{\sn Offers Fast Scalable HTAP Queries.}
Figure \ref{fig.eva.chbenchmark} shows the response time of the representative analytical queries (Q1 and Q6) at different concurrency levels (with 8 or 16 \emph{Transactional} and 1 or 4 \emph{Analytical} threads).
Bw-Tree and ART show similar performance trends with \bt in our evaluation, so we denote them as \emph{BTree}.
We make the following observations.
(1) For Q1, which has high selectivity (36\%) and mainly performs aggregation, BTree indexing hurts analytical queries.
However, \sn is comparable to Scan.
(2) For Q6, which has low selectivity (1.8\%) and spends more time scanning tuples, BTree indexing outperforms the scan.
\sn further reduces the response time and achieves 2--11$\times$ performance improvement compared to the baselines, irrespective of concurrent TPC-C updates. 
(3) For both Q1 and Q6, increasing the number of analytical threads from 1 to 4 does not noticeably slow down \sns, 
but it hurts scan and increases the response time by $\sim$3$\times$
mainly because the cache miss rate increases by $\sim$2$\times$ (reported by Linux perf in profiling \emph{capacity misses}).
In addition, increasing the number of TPC-C threads from 8 to 16 incurs higher contention and leads to increased query latency for BTree, while CUBIT is barely affected.
Overall, \sn is a promising indexing candidate for analytical queries on HTAP workloads.

\vspace{-0.06in}
\section{Related Work}\label{sec.related}
\vspace{-0.02in}


\Paragraph{Offering Faster Queries.}
The majority of research (e.g., TEB~\cite{Lang2020} and BinDex~\cite{Li2020a}) focus on improving query performance by leveraging customized data structures that are read-only.
Although researchers have been improving their updatability \cite{Saputra2022}, the updates remain costly \cite{Li2020a}.
\sn inherently offers efficient updates and is orthogonal to these designs.

\Paragraph{Compressing Bitvectors.}
Run-length encoding and its variants \cite{Antoshenkov1995, Wu2006, Colantonio2010, Lemire2018} have been widely used to compress bitvectors.
The latest research, Roaring~\cite{Lemire2018}, splits a bitvector into fixed-length containers that are recorded and compressed independently.
These designs support updates in the granularity of bitvector (or container), which suffer from performance deterioration when naively used in updatable bitmap indexes.
However, \sn provides a framework to atomically and efficiently update multiple bitvectors, and thus is orthogonal to these designs.
We select the RLE-based WAH~\cite{Wu2006} in \sns's implementation and leave utilizing Roaring, which contains the structure modification operations (i.e., changing an array container to a bitmap one, and vice versa), as future work.

\Paragraph{Bit-Packing.}
Bitmap indexes and bit-packed storage (e.g., SIMD-Scan \cite{Willhalm2009}, BitWeaving \cite{Li2013}, and ByteSlice \cite{Feng2015}) are closely related regarding space efficiency and fast data retrieval.
Nevertheless, bit-packed storage is primarily optimized for fast scanning, making merging results from multiple scans typically expensive \cite{Li2013}.
Another trend involves incorporating lightweight (e.g., Column Imprints \cite{Sidirourgos2013} and Column Sketches \cite{Hentschel2018}) and customized indexes (e.g., BinDex \cite{Li2020a} and Cabin \cite{Chen2024}) to accelerate scans.
These research results highlight the necessity of maintaining indexes on bit-packed storage, and \sn is an effort toward a general solution. 

\Paragraph{Modeling Analysis.}
Prior research \cite{ONeil1997, Chan1998, Chan1999} has extensively examined different bitmap indexes in terms of space-time tradeoffs.
However, existing conclusions may not fully apply to \sn that focuses on updates.
To more effectively analyze \sns, a new model that considers the specific characteristics of updates is necessary, which we leave as our future work.




\vspace{-0.06in}
\section{Conclusion}\label{sec.conclusion}
\vspace{-0.02in}

We present \sns, a bitmap index that supports scalable, real-time updates, enabling a departure from the traditional approaches of updating bitmap indexes which are notably in a serial, batch mode.
\sn expands the applicability of bitmap indexing by, for the first time, enabling the efficient maintenance of bitmap indexes on frequently updated attributes, thereby demonstrating the potential to accelerate a DBMS's critical operations including Scan, Join, and Aggregation.
Experimenting with OLAP and HTAP workloads demonstrates that the \sns-powered query engine delivers remarkable performance improvement for analytical queries.

\balance
\clearpage

\bibliographystyle{ACM-Reference-Format}
\bibliography{library-manos-mendeley,library-urls}

\clearpage
\appendix

\section*{Appendix to the \sn paper}\label{sec.appendix}


\section{Programming Semantics of \sns}\label{sec.appendix.semantics} 

\sn conforms to the traditional specification for secondary indexes in DBMSs and provides an API that supports \vn{Query}, \vn{Update}, \vn{Delete}, and \vn{Insert} operations.

\begin{itemize}[leftmargin=1.3em]

    \item \vn{Query(value or range)} accepts a predicate in the form of a pointer of range query and returns the result as either a bitvector or an array of pointers to the matching tuples.

    \item \vn{Update(row, val)} retrieves the current value of the specified \emph{row}, updates it to \emph{val}, and returns \emph{true} if the operation succeeds.

    \item \vn{Delete(row)} retrieves the current value of the specified \emph{row}, deletes the row, and returns \emph{true} if the operation succeeds.

    \item \vn{Insert(val)} appends the value \emph{val} to the tail of the bitmap index, increments global variables like \emph{N\_ROWS}, and returns \emph{true} if the insertion succeeds.

\end{itemize}


\section{Latch-free \sns}\label{sec.appendix.lockfree}

\Paragraph{Concurrency Issues with \sns-lk.}
\sns-lk employs a read-write latch to serialize concurrent UDIs that attempt to append \emph{ULEs} at the tail of the per-index Delta Log.
This design stems from the classic MVCC mechanisms, where concurrent updates use latches to prevent simultaneous modifications to the same portion of data (e.g., tuples or pages).
Experimentally, we found that this latch can lead to long-tail UDI latency due to two main factors:
(1) UDIs on bitmap indexes lock at the granularity of bitvectors, which are typically far fewer in number than the tuples or pages locked by traditional MVCC mechanisms,
and (2) skewed UDIs concentrate on a few hotspot bitvectors, resulting in even higher contention.
Additionally, \sn faces severe time constraints inherent in indexing in DBMSs.

\Paragraph{Solution.}
We address the above-described challenges from two angles.
First, when UDIs and merge operations conflict (i.e., both attempt to append their \emph{ULEs} to the tail of the Delta Log simultaneously), they consolidate their \emph{ULEs} and delegate committing them to subsequent UDIs.
Second, instead of busy-waiting, suspended UDIs \emph{help} the other in making progress until completion, and then retry.
The resulting algorithm, termed \sns-lf, offers non-blocking (latch-free) UDIs that never block one another, ensuring the system always makes progress without suspension or deadlock.
This approach effectively eliminates the primary cause of unexpectedly long tail latency of UDIs associated with MVCC.

\SubParagraph{Helping Mechanism Basis.}
Under the hood, we employ Michael and Scott's seminal lock-free first-in-first-out linked list~\cite{Michael1996}, known as the \emph{MS-Queue}, for the implementation of Delta Log.
The MS-Queue facilitates latch-free insert and delete operations that do not block one another.

Specifically, an insert operation $A$ assigns the \emph{next} pointer of the list's last node to a new node $n$, by using an atomic \emph{compare-and-swap (CAS)} instruction.
A successfully execution of this CAS operation marks the linearization point of $A$, implying that \emph{n} has been successfully inserted into the list with respect to other concurrent operations.
Subsequently, the operation $A$ attempts to update the global TAIL pointer to reference \emph{n} through another \emph{CAS} instruction.

Should $A$ be suspended before performing the second \emph{CAS}, a concurrent insert operation $B$, which failed because of the successful insertion of the node $n$, will first \emph{help} $A$ by updating the global pointer TAIL by also using \emph{CAS} instructions. 
Following this assistance, operation $B$ will restart its process.
Delete operations synchronize in a similar fashion.

\SubParagraph{Challenges.}
In the context of the MS-Queue, assisting operation $A$ merely requires operation $B$ to update the TAIL pointer.
However, in the case of \sns, a UDI must update several global variables, including TAIL, TIMESTAMP, and/or N\_ROWS, and a merge operation is required to update TIMESTAMP, TAIL, and the head pointer of the corresponding version chain.
We thus extend the standard helping mechanism to accommodate these additional requirements.

\SubParagraph{Our Helping Mechanism for \sns.}
We propose a \emph{helping} mechanism designed to atomically update a group of variables, drawing inspiration from recent latch-free designs~\cite{Arbel-Raviv2018}.
Specifically, each UDI and merge operation records the old and new values of the variables to be updated in its \emph{ULE} before appending it to the tail of Delta Log by using a \emph{CAS} instruction.
Once this step $O_1$ succeeds (i.e., this UDI operation linearizes), the \emph{ULE} becomes accessible to other threads via the next pointers of the \emph{ULEs} in Delta Log.
Should another UDI or merge operation $O_2$ fail to append its \emph{ULE}, it first assists $O_1$ in completion.
Specifically, for each variable to be updated, $O_2$ retrieves the old and new values, and updates the variable to its new value by using \emph{CAS} instructions.
If a \emph{CAS} fails, indicating that this variable has already been updated by either $O_1$ or other assisting operations, $O_2$ simply skips updating this variable.
After helping update all variables in $O_1$'s \emph{ULE}, $O_2$ starts over.

\Paragraph{Correctness.}
\sns-lf is correct, which we prove from the following aspects.

\SubParagraph{Immune to ABA problem.}
In theory, the ABA problem may occur when updating the TIMESTAMP and N\_ROWS variables.
However, it would take TIMESTAMP more than one million years to wraparound under a scenario with 500K UDIs per second. 
Similarly, N\_ROWS is a 64-bit that monotonically increases, making the likelihood of an ABA problem extremely low.
Additionally, the ABA problem is entirely avoided for other variables (e.g., TAIL and the pointers to the version chains) due to the epoch-based reclamation mechanism employed in \sn, which ensures that no memory space is reclaimed and reused while any worker thread still holds a reference to it.
Therefore, we conclude that, in practice, \sns-lf is effectively immune to the ABA problem.

\SubParagraph{No-Bad-Thing-Happen (Correctness) Property.}
We define the term \emph{shared variables} as the global variables that are updated by UDI and merge operations.
The correctness of \sns-lf is ensured by the following key facts.
(1) Shared variables can only be updated after a \emph{ULE} has been successfully appended to the tail of Delta Log.
(2) The manner in which shared variables are updated is pre-defined in the \emph{ULE} by specifying both the old and new values for each variable.
(3) The updating of shared variables can be performed by any active threads, allowing concurrent threads to help one another.
(4) Shared variables are updated exclusively using \emph{CAS} instructions.
(5) The ABA problem is effectively mitigated.
Overall, \sns-lf guarantees that when a UDI and merge operation associated with a \emph{ULE} completes, each shared variable (a) has been updated to the specified new value, and (b) has been updated only once.

\SubParagraph{Good-Thing-Always-Happen (Liveness) Property.}
The arguments on linearization points (as discussed above) indicate that \sns-lf provides wait-free queries and latch-free UDIs, meaning that these operations never experience blocking.


\section{Size of HUDs}\label{sec.appendix.hud}
In general, a HUD initially contains only 0s.
As UDIs accumulate, the number of 1s in a HUD increases, so does the size of the HUD (stored compactly as a list of positions).
We now study the operation sequences that increase the size of a HUD, and then show that it is very unlikely that a HUB contains more than two positions.

\begin{figure}[t]
    \includegraphics[width=0.9\columnwidth]{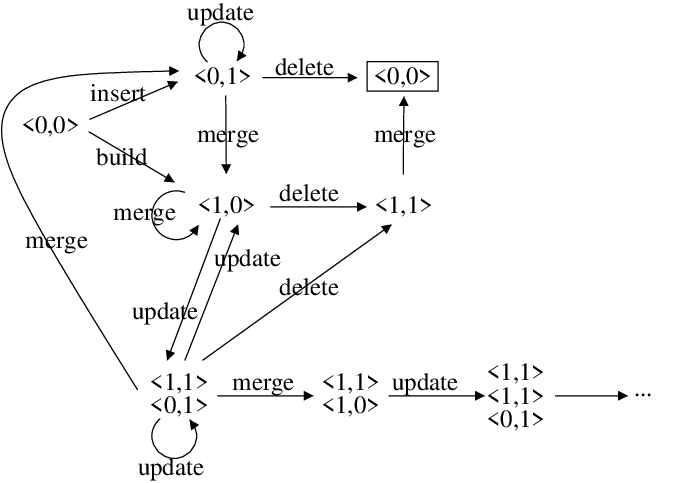}
    \caption{Finite-State Machine of the HUDs of a row by only recording the <$U_i$, $V_i$> pairs that contain 1s.
            Except for the bottom-right state, the number of 1s recorded in a HUD is zero to two.
            The only sequence of operations that increases the number of 1s in HUDs is
            a sequence of interleaved update and merge operations, which takes place with low probability.}
    \label{correct.transition}
\end{figure}

\Paragraph{FSM of HUDs.}
Conceptually, a HUD is a bit-array with a length equal to the cardinality of the domain, and the $i^{th}$ bit in this array, denoted $U_i$, is associated with the corresponding bit of the $i^{th}$ VB, denoted $V_i$.
We study the transition of the HUD by using a Finite-State Machine (FSM), in which each node records the <$U_i$, $V_i$> pairs for all possible $i$, denoted <$U$, $V$>$_i$.
For ease of presentation, except for the initial state (the top-left node indicating that the row is just allocated) and the final state (the top-right node indicating that the row has been deleted),
all the <0, 0> pairs are removed.
Each arrow is labeled with the operation that triggers the transition.
For example, an insert operation allocates a new HUD and changes its state from <0, 0> to <0, 1>, indicating that the corresponding bit of the HUD has been set to 1.
An update may change a HUD from '<0,0>,<0,1>' to '<0,1>,<0,0>', leading to a circular arrow starting from and ending at the same node.
That is, there is no transition to a new state because the <0, 0> pairs are omitted in the FSM.
The complete FSM is shown in Figure \ref{correct.transition}. We make the following observations.

(1) Except for the bottom-right state, the number of 1s in each HUD is zero to two with high probability.

(2) The only operation sequence that increases the number of 1s of a HUD (assume <0, 1>$_{i1}$ initially) is as follows.

\begin{itemize}

    \item \textbf{A1}: A merge happens on the VB $i_1$, resulting in the state <1, 0>$_{i1}$.

    \item \textbf{A2}: An update changes this row to value $i_2$, resulting in the state <1, 1>$_{i1}$<0, 1>$_{i2}$.

    \item \textbf{A3}: A merge happens on the VB $i_2$, resulting in the state <1, 1>$_{i1}$<1, 0>$_{i2}$.

    \item \textbf{A4}: A subsequent update changes this row to any values except $i_1$, denoted as $i_3$, resulting in a HUD with three 1s: <1, 1>$_{i1}$<1, 1>$_{i2}$<0, 1>$_{i3}$. 

    \item This resulting HUD is  <R, 3, 1, 2, 3>.
\end{itemize}

In summary, if (a) updates always change a row to new values, (b) update and merge operations happen alternatively and each merge always happens on the new value of the preceding update, and (c) no deletes happen, the number of 1s in a HUD can grow.
Assume that there is no delete and that updates and merges happen uniformly on all possible values.
The probability of A1 is $1/c$, where $c$ is the cardinality,
the probability of A2 is $(c-1)/c$, and
the probability of A3 and A4 are $1/c$ and $(c-2)/c$, respectively.
Overall, a HUD will contain $n$ 1s with a probability less than $1/c^{n-1}$.
For example, when $c = 128$, the probability that a HUD contains seven 1s is $1/128^{6} = 1/2^{42}$, which happens extremely unlikely.


\section{Existing Updatable Bitmap Indexes}

\Paragraph{In-place.}
The most straightforward approach, denoted \emph{In-place} \cite{Silberschatz2020}, directly updates the underlying bit-matrix.
In order to update the $k^{th}$ row from value $v_1$ to $v_2$,
\emph{In-place} applies the \emph{decode-flip-encode} procedure on both bitvectors of $v_1$ and $v_2$.
To delete the $k^{th}$ row, \emph{In-place} applies the same procedure on the bitvector of the old value and sets its $k^{th}$ bit from 1 to 0.
An insert operation appends bit 1 at the tail of the bitvector of the corresponding value, and then appends bit 0 to the others.
\emph{In-place}'s inferior performance comes from the time-consuming decode-flip-encode procedure.

\Paragraph{UCB.}
To alleviate the performance issue of In-place, UCB \cite{Canahuate2007} introduces an extra bitvector, denoted \emph{existence bitvector} (EB), that indicates whether a given row is valid or not.
Initially, all bits in EB are 1s. 
A delete operation is performed by setting the corresponding bit in EB to 0.
An insert operation appends a 1 to the tail of EB, and increments the global variable \emph{N\_ROWS} that indicates the number of rows in UCB.
An update operation is transformed to delete-then-append operations; that is, the new value is appended at the tail of the bitvector, and a mapping between the invalidated row ID and the new-appended row ID is kept.
By avoiding decoding and then encoding the value bitvectors, UDIs of UCB are supposed to be more efficient than In-place.
The efficiency of UCB is predicated on EB being highly compressible.
In practice, however, its performance deteriorates sharply as the total number of UDIs performed increases and EB becomes less compressible \cite{Athanassoulis2016a}.
Meanwhile, UCB does not provide a standard \emph{set} abstract data type,
and programmers must maintain an additional level of indirection to map each row ID from users' perspective to UCB's, which incurs considerable performance penalties as the number of updates increases.

\Paragraph{UpBit.}
To address the above-discussed issues, the state-of-the-art solution, \ubs~\cite{Athanassoulis2016a}, 
associates an additional \emph{update bitvector} (UB) with each \emph{value bitvector} (VB) in the domain of the indexed attribute.
UBs keep track of updates to VBs; that is, UDIs flip bits in UBs that are merged back to VBs in a lazy and batch manner.
Therefore, UBs are highly compressible, resulting in reduced decode-flip-encode overheads.


\section{Parallelizing Bitmap Indexes}\label{sec.appendix.parallelize}
\Paragraph{\ubs.}
We parallelize \ub, the state-of-the-art updatable bitmap index, by using a fine-grained locking mechanism.
Specifically, the <VB, UB> pair of every value $v$ is protected by a reader-writer latch, denoted $latch_{v}$.
Global variables like \emph{N\_ROWS} are protected by a global latch $latch_{g}$.
Update and delete operations first acquire the $latch_{v}$ of all values in shared mode to retrieve the current value of the specified row. 
Then, they upgrade $latch_{v}$ of the corresponding bitvectors to exclusive mode in order to flip the necessary bits.
An insert operation acquires $latch_{g}$ and the corresponding $latch_{v}$ in exclusive mode.
Consequently, a query operation acquires $latch_{g}$ and the corresponding \vn{$latch_{v}$} in shared mode.

\Paragraph{UCB.} UCB's UDIs update the only EB, and queries read this EB simultaneously.
Therefore, we parallelize UCB by using a global reader-writer latch to serialize concurrent queries and UDIs.
An insert operation holds this latch before updating the global variable \emph{N\_ROWS}.

\Paragraph{In-place.} One way to parallelize In-place is to use fine-grained reader-writer latches, the same as in \ubs.
However, with this mechanism, an insert operation needs to acquire $cardinality$ latches before appending bits to the tail of all the VBs, dramatically reducing the overall throughput.
Therefore, we parallelize In-place by using a global reader-writer latch, the same as for UCB.
Surprisingly, the parallelized In-place outperforms UCB for high concurrency (see the evaluation results in our paper).


\section{TPC-H}\label{sec.appendix.tpch}
We integrated \sn into DuckDB and implemented 12 of the 22 TPC-H queries, along with the New Sales Refresh Function (RF1) and Old Sales Refresh Function (RF2) transactions.
We optimized as many choke points in DuckDB's query engine as possible \cite{Boncz2013, Dreseler2020}.
For example, in TPC-H Q1, we represented the group-by expression as small integers within a narrow range (instead of using a String class) and utilized an array (rather than a hash table) to store aggregation statistics, which improved query performance by 7\%.
By excluding certain group-by attributes that can be derived from the primary key, we achieved a 46\% performance improvement in Q10.
We optimize its Scan operator by (1) storing column values more compactly (e.g., by compressing \emph{l\_quantity} from 13 to 6 bits) with a storage layout inspired by BitWeaving's HBP \cite{Li2013}, and (2) scanning columns using SIMD instructions, leading to a 36\% lower latency for Q6.
We refer to our optimized version as DuckDB$^+$ and use it as the baseline.

In our experiments, we set the Scale Factor (SF) of our workload to 10.
\emph{Note that in this section, we present the pseudocode for TPC-H queries for reader's reference; they are identical to those listed in the TPC-H specification.}

\begin{algorithm2e}[h]
    \footnotesize
    \AlgoDisplayBlockMarkers
    \SetAlgoNoLine

    \textbf{LOOP} 1,500 times\\
    \quad \textbf{INSERT} a new row into the ORDERS table\\
    \quad \textbf{LOOP} random[1, 7] times\\
        \quad \quad \textbf{INSERT} a new row into the LINEITEM table\\
    \quad \textbf{END LOOP}\\
    \textbf{END LOOP}

    \caption{TPC-H RF1.}
    \label{alg.tpch.RF1}
\end{algorithm2e}

\begin{algorithm2e}[h]
    \footnotesize
    \AlgoDisplayBlockMarkers
    \SetAlgoNoLine

    \textbf{LOOP} 1,500 times\\
    \quad \textbf{DELETE} from ORDERS where o\_orderkey = [VALUE]\\
    \quad \textbf{DELETE} from LINEITEM where l\_orderkey = [VALUE]\\
    \textbf{END LOOP}

    \caption{TPC-H RF2.}
    \label{alg.tpch.RF2}
\end{algorithm2e}

\Paragraph{Refresh Workloads.}
The pseudocode for RF1 and RF2 is presented in Algorithms \ref{alg.tpch.RF1} and \ref{alg.tpch.RF2}.
These algorithms update the LINEITEM and ORDERS tables, along with the associated indexes.
Due to their update-friendly nature, \sn instances on the attributes of these two tables do not incur any maintenance downtime when RF1 and RF2 are executed in our experiments.
In contrast, with other bitmap indexes, RF1 and RF2 must be performed during a scheduled maintenance window, during which indexes are unavailable.

A dedicated worker thread, referred to as the \emph{RF Thread}, is assigned to execute RF1 and RF2 periodically, while other worker threads concurrently perform query transactions.
During each execution, the RF thread invokes RF1 or RF2, modifying 1,500 tuples in the ORDERS table and $\sim$4,500 tuples in the LINEITEM table, followed by a batch update of the corresponding indexes.
After each execution, the RF thread waits until the ratio of the overall workload of queries to refreshes reaches 98:2, before initiating the next RF transaction.

\section{CH-benCHmark}\label{sec.appendix.chbench}
We implemented the CH-benCHmark~\cite{Cole2011} that consists of a full version of the TPC-C benchmark and a set of TPC-H-equivalent analytical queries on the same tables.

\Paragraph{Selectivity.}
In our evaluation, we found that many attributes in CH-benCHmark cover a narrow scope, such that the queries unreasonably select almost all of the tuples.
We thus modified the propagated values and the query predicates to provide a reasonable selectivity.
For example, we set the values of the \emph{ol\_delivery\_d} attribute in the \emph{ORDER-LINE} table in the range of [1983, 2023), and the values of the \emph{ol\_quantity} attribute in the range of [1, 25000), both in a uniform distribution.
As a consequence, each CH-benCHmark Q1 selects rows on years (16 out of 40) and delivery state (9 out of 10), leading to an average selectivity of $\frac{16}{40}\times\frac{9}{10} \approx 36\%$,
and each Q6 selects rows on years (20 out of 40), quantities (1000 out of 25,000), and delivery state (9 out of 10), leading to an average selectivity of $\frac{20}{40}\times\frac{1}{25}\times\frac{9}{10} \approx 1.8\%$.
The SQL code for the Q1 and Q6 of CH-benCHmark are listed in Algorithms \ref{alg.chbench.Q1} and \ref{alg.chbench.Q6}.

\begin{algorithm2e}[h]
    \footnotesize
    \AlgoDisplayBlockMarkers
    \SetAlgoNoLine

    \textbf{SELECT} ol\_number,\\
                \qquad \qquad sum(ol\_quantity) as sum\_qty,\\
                \qquad \qquad sum(ol\_amount) as sum\_amount,\\
                \qquad \qquad avg(ol\_quantity) as avg\_qty,\\
                \qquad \qquad avg(ol\_amount) as avg\_amount,\\
                \qquad \qquad count($\ast$) as count\_order\\
    \textbf{FROM} \quad \quad orderline\\
    \textbf{WHERE} \quad \quad ol\_delivery\_d > `2007-01-02 00:00:00.000000'\\
    \textbf{GROUP BY} \quad ol\_number \textbf{ORDER BY} ol\_number

    \caption{CH-benCHmark Q1.}
    \label{alg.chbench.Q1}
\end{algorithm2e}

\begin{algorithm2e}[h]
    \footnotesize
    \AlgoDisplayBlockMarkers
    \SetAlgoNoLine

    \textbf{SELECT} sum(ol\_amount) as revenue\\
    \textbf{FROM}  orderline\\
    \textbf{WHERE}  ol\_delivery\_d >= `1999-01-01 00:00:00.000000'\\
                    \qquad \qquad \textbf{and} ol\_delivery\_d < `2020-01-01 00:00:00.000000'\\
                    \qquad \qquad \textbf{and} ol\_quantity between 1 and 1000

    \caption{CH-benCHmark Q6.}
    \label{alg.chbench.Q6}
\end{algorithm2e}

\section{Impact of Data Skew}\label{sec.eva.zipf}

The distribution of data among bitvectors plays a key role in performance for two reasons.
First, biased distributions may lead to few \emph{target} bitvectors containing many more 1s than others, making them less compressible.
Second, the target bitvectors face higher contention levels among concurrent UDIs. 
We thus evaluate the impact of a skewed distribution.
We use the same configuration with the highest concurrency level (32 worker threads). The dataset follows the Zipfian distribution with the skew parameter $\alpha$ being set to 1.5, which implies that about 40\% of the entries have the two most popular values, and the remaining are uniformly distributed in the entire value domain.
We make the following observations.

\Paragraph{All Bitmap Indexes Have 2$\times$ Faster Queries for Skewed Data.}
The overall throughput and mean query latency of \textit{all indexes} are improved by about 2$\times$, compared to the case with uniform data distribution.
\textit{The reason is that most bitvectors contain few 1s, and are thus highly compressible.}
Queries on these bitvectors are very fast.

\begin{figure}[t]
    \centering
    \vspace{-0.16in}
    \setlength{\belowcaptionskip}{-10pt}
    \hspace*{-0.8em}
    \subfloat {
        \includegraphics[scale=0.2]{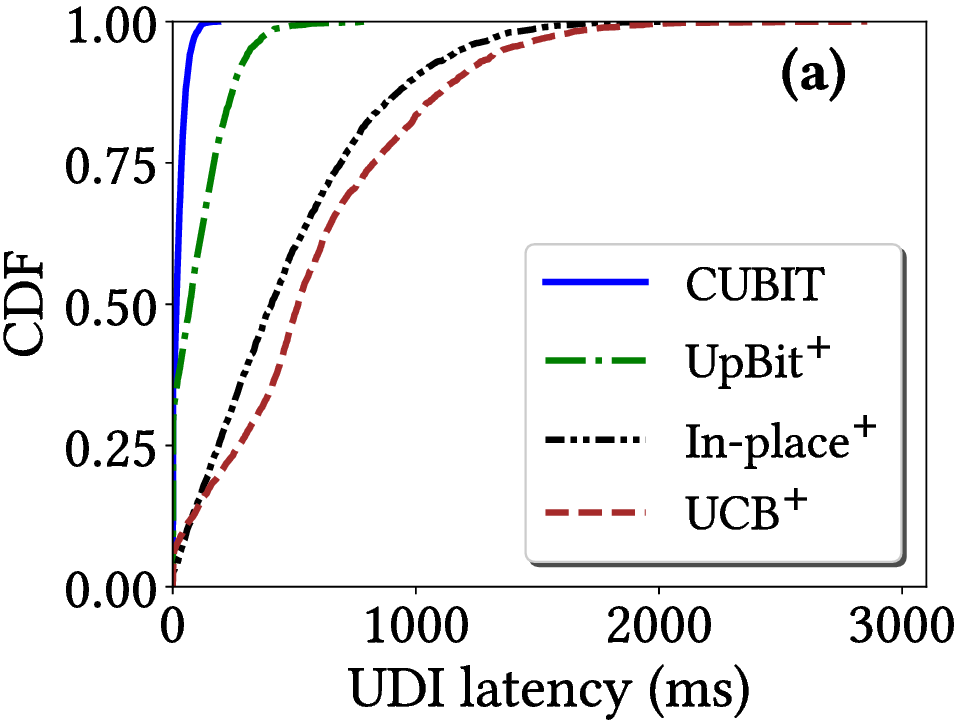}
        \label{eva.zipf.udi}      
    }
    \hspace*{-0.1em}
    \subfloat {
        \scriptsize
        \begin{tabular}[b]{lrrr}
        \hline
        Index Approach & Median & P99 & P99999 \\
        \hline
    	\sns-lf-unif	& 11		&  68		& 152		\\
    	\sns-lf-zipf	& 15		&  76   	& \textbf{153}		\\
        \hline
    	\sns-unif	& 11		&  71		& 127		\\
    	\sns-zipf	& 15		& 111		& \textbf{185}		\\
        \hline
    	UpBit-unif		& 52		& 430	    & 656   	\\
    	UpBit-zipf		&  71		& 457	    & \textbf{785}   	\\
        \hline
    	In-place-unif    & 214   	& 1,008 	& 1,882 	\\
    	In-place-zipf    & 407   	& 1,523 	& \textbf{2,009} 	\\
        \hline
        \end{tabular}
        \label{eva.zipf.details}      
    }
    \vspace{-0.1in}
    \caption{(a) \sn's UDI latency is not affected by data skew (Zipfian, $\alpha$$=$$1.5$), and (table) has superior tail latency.}
    \label{eva.zipf}      
    \vspace{-0.01in}
\end{figure}

\Paragraph{\sn Has Stable UDI Latency for Skewed Data.}
For all approaches, UDI latency increases for skewed data because 40\% of the UDIs involve a few bitvectors, leading to high contention on them.
However, \sn remains the most stable design (as shown in Figure \ref{eva.zipf.udi}).
Note that UCB$^{+}$ performance deterioration depends on the total number of UDIs performed; thus, we omit it in the remainder of the analysis.

To better understand the tail UDI latency, we zoom in on Figure~\ref{eva.zipf.udi} and list the corresponding statistics (\textit{-zipf}) in Figure \ref{eva.zipf.details}.
We also list the statistics of the same experiments with uniform distributions (\textit{-unif}) for comparison.
We make the following observations.
(A) First, \textit{\ubs$^{+}$ significantly reduces UDI's tail latency when compared to In-place$^{+}$ because of its fine-grained locking mechanism.}
Figure~\ref{eva.zipf.details} shows that the P99999 UDI latency of \ubs$^{+}$ is 2.6$\times$ smaller than that of In-place$^{+}$.
(B) Second, \textit{the fine-grained read-write locking mechanism employed by \ubs$^{+}$ cannot alleviate the contention on the target bitvectors.
In contrast, \sn's lightweight UDIs reduce contention among hot bitvectors.}
For example, the P99999 latency of basic \sn is about 4.2$\times$ smaller than that of \ubs$^{+}$.
(C) Third, the \textit{helping mechanism} employed by \sns-lf reduces the number of latches acquired, further reducing \sn's P99 UDI latency by 31\% and P99999 UDI latency by 17\%.

\section{\sn for OLTP}\label{sec.eva.oltp}

\Paragraph{TPC-C.}
We reuse the DBx1000 framework to implement a full-blown TPC-C benchmark to test \sn for a pure transactional workload.
We compare the performance of Stock-Level (SL) transactions with a \bt and \sn on the \emph{Quantity} attribute. 
In our evaluation, \emph{\#Warehouse} = 100, and the \emph{Stock} table contains about 10M tuples.
We make the following two observations.


\Paragraph{Update Friendly.}
Despite that other transactions in TPC-C (e.g., \emph{New-Order}) heavily update the \emph{Quantity} attribute and the associated \sn indexes,
\sns-assisted SL is competitive to other indexes, demonstrating that \sn is update-friendly and does not introduce noticeable maintenance overhead.

\Paragraph{\sn Brings Limited Performance Gains to OLTP.} 
\sns-powered DBMS reduces the response time of SL from 0.55ms to 0.54ms by assigning 8 cores to each query.
The performance gain is mainly because querying \sn is easier to parallelize.
However, the improvement is not noticeable because queries in OLTP are very selective (e.g., each SL yields about 200 matching entries), such that other indexes like \bt also perform well.
Overall, our conclusion is that \sn can be used in OLTP DBMSs without incurring performance penalties, 
but it is a better fit for OLAP and HTAP with inherent moderate selectivity.

\end{document}